%% file: ApJ-DFDItheory.tex
\documentclass[preprint]{aastex}

\usepackage{amsmath,natbib,url}
\input{usersetcommands}

\shorttitle{Theory of DFDI}
\shortauthors{van~Eyken, Ge, \& Mahadevan}

\begin{document}


\title{Theory of Dispersed Fixed-Delay Interferometry for Radial Velocity
  Exoplanet Searches}


\author{Julian~C.~van Eyken\altaffilmark{1,2,5}, Jian~Ge\altaffilmark{1,5}, and Suvrath~Mahadevan\altaffilmark{1,3,4,5}}
\email{vaneyken@ipac.caltech.edu, jge@astro.ufl.edu, suvrath@astro.psu.edu}

\altaffiltext{1}{Department of Astronomy, University of Florida, 211 Bryant
  Space Science Center, PO~Box~112055, Gainesville, FL, 32611-2055,
  USA}
\altaffiltext{2}{NASA Exoplanet Science Institute,
  California Institute of Technology, 770 South Wilson Avenue,
  M/S~100-22, Pasadena, CA, 91125, USA}
\altaffiltext{3}{Department of Astronomy \&
  Astrophysics, The Pennsylvania State University, University Park,
  PA 16802, USA}
\altaffiltext{4}{Center for Exoplanets and Habitable Worlds, The
  Pennsylvania State University, University Park, PA 16802, USA}
\altaffiltext{5}{Visiting Astronomers, Kitt Peak National Observatory,
  National Optical Astronomy Observatory, which is operated by the
  Association of Universities for Research in Astronomy, Inc. (AURA),
  under cooperative agreement with the National Science Foundation}



\begin{abstract}
\input{abstract}
\end{abstract}


\keywords{instrumentation: interferometers --- instrumentation:
  spectrographs --- methods: analytical --- planetary systems ---
  techniques: radial
  velocities}



\input{section1}

\input{section2}

\input{section3}

\input{section4}

\acknowledgments
\input{acknowledgments}

\appendix
\input{appendixA}

\input{appendixB}
\input{appendixC}

\bibliographystyle{apj} 
\bibliography{juliansbibliography}

\clearpage

\end{document}

%% file: usersetcommands.tex
\newcommand{\kms}{\,{\rm km}\,{\rm s}^{-1}}
\newcommand{\ms}{\,{\rm m\,s^{-1}}}

\newcommand{\mm}{\,{\rm mm}}
\newcommand{\msrad}{\,{\rm m\, s^{-1}\, rad^{-1}}}

\newcommand{\ang}{\,\text{\AA}}
\newcommand{\ii}{\mathrm{i}}
\newcommand{\dd}{\,\mathrm{d}}
\newcommand{\FT}{\mathcal{F}}
\newcommand{\ee}{\mathrm{e}}

\newcommand{\m}{\,{\rm m}}

\newcommand{\Mj}{\, M_{\rm J}}

\newcommand{\vmag}{\,{\rm mag}}

%% file: abstract.tex
The dispersed fixed-delay interferometer (DFDI) represents a new
instrument concept for high-precision radial velocity (RV) surveys for
extrasolar planets. A combination of Michelson interferometer and
medium-resolution spectrograph, it has the potential for performing
multi-object surveys, where most previous RV techniques
have been limited to observing only one target at a time. Because of
the large sample of extrasolar planets needed to better understand
planetary formation, evolution, and prevalence, this new technique
represents a logical next step in instrumentation for RV
extrasolar planet searches, and has been proven with the single-object
Exoplanet Tracker (ET) at Kitt Peak National Observatory, and the
multi-object W.~M.~Keck/MARVELS Exoplanet Tracker at Apache Point
Observatory. The development of the ET instruments has necessitated
fleshing out a detailed understanding of the physical principles of
the DFDI technique. Here we summarize the fundamental theoretical
material needed to understand the technique and provide an overview of
the physics underlying the instrument's working. We also derive some
useful analytical formulae that can be used to estimate the level of
various sources of error generic to the technique, such as photon shot
noise when using a fiducial reference spectrum, contamination by
secondary spectra (e.g., crowded sources, spectroscopic binaries, or
moonlight contamination), residual interferometer comb, and reference
cross-talk error. Following this, we show that the use of a
traditional gas absorption fiducial reference with a DFDI can incur
significant systematic errors that must be taken into account at the
precision levels required to detect extrasolar planets.

%% file: section1.tex
\section{The DFDI Concept and the ET Program}

\subsection{The Need for a New Instrument}\label{theneed}

Despite the remarkable achievements in extrasolar planet detection
over the last decade, identification of many more planets is still
needed to constrain formation and evolutionary models. This is
partially because of the unexpected diversity of planet properties
uncovered, and partially because of a lack of large, well-defined,
unbiased target search lists -- the primary concern naturally having
been to find planets in the first place. To this point many surveys
have been subject to completeness issues or in some cases deliberate
biases toward planet detection \citep[e.g.][]{elodieFeHbiased}, making
it difficult to perform robust statistical analyses of the known
planet sample. \citet{Armitage} concluded that there is still a strong
need for large uniform surveys to enlarge the statistical sample
available: drawing on the unbiased survey of \citet{FischerValenti05},
he was only able to find a uniform subsample of 22 of the over 170
planets then known that satisfied the requirements for a statistical
comparison with models.

A few thousand stars have been searched between the various RV surveys
since the first RV discoveries of giant extrasolar planets around
solar-type stars \citep{51PegDiscovery}, including a large fraction of
the late-type, stable stars down to visual magnitude
$\sim8$. Improved instrument light throughput would help facilitate the
survey of fainter stars. \citep[A review of radial velocity (RV) discoveries is given
by][] {Udry07}. Although the rate of detections from transit
surveys will likely increase, transit surveys can only detect the
small fraction of planets which happen to eclipse their parent stars
\citep[$\sim10\%$ probability for hot Jupiters, from geometrical
considerations --][]{WASP}. Furthermore, the complementary information
gained from RV detections remains of great value. There is therefore a
strong case for finding a technique capable of RV surveys down to
faint magnitudes and at faster speeds than have been achieved over the
last decade. The Exoplanet Tracker (ET) instruments are a new type of
fiber-fed radial velocity (RV) instrument based on the `dispersed
fixed-delay interferometer' (DFDI), built with the goal of satisfying
this requirement.

\subsection{The DFDI Principle}

The radial velocity technique for detecting exoplanets consists in
measuring the reflex motion of the parent star due to an orbiting
planet by measuring very precisely the resulting Doppler shifts of the
stellar absorption lines. Achieving this requires extremely high
precision: internal precisions now typically reach down to the $3\ms$
level \citep{Butler3ms,Vogt00}, and even as low as $1\ms$ or better
\citep{harps2}. For comparison, a Jupiter analogue in a circular
orbit around a solar-type star would cause sinusoidal radial velocity
variations with an amplitude of about $12.5\ms$. Exoplanet radial
velocity surveys have traditionally depended on recording very high
resolution echelle spectra, either cross correlating the spectra with
reference template spectra, or fitting functions to the line profiles
themselves to measure the positions of the centroids.

The DFDI technique, upon which the Exoplanet Tracker (ET) instruments
are based, comprises a Michelson interferometer followed by a low or
medium resolution post-disperser (also referred to by
\citet{Erskine03} as an externally dispersed interferometer, or `EDI',
emphasizing the distinction from techniques where the dispersing
element is internal to the interferometer). The effective resolution
of the instrument is determined primarily by the interferometer, so
the post-dispersing spectrograph can be of much lower resolution than
in traditional dispersive techniques, and consequently can be smaller,
cheaper, and have higher throughput
\citep{Ge2002,Ge03ASP,Ge03SPIE4838}. The technique is closely related
to Fourier transform spectroscopy: the post-disperser effectively
creates a continuum of very narrow bandpasses for the interferometer,
increasing the interference fringe contrast. All the information
needed is contained in the fringe phase and visibility. It emerges
that since we are only interested in the Doppler shift of the lines,
measurements are required at essentially only one value of
interferometer delay (hence `fixed delay').

The cost of the instrument is comparatively low, and most importantly,
it can operate in a single-order mode: where traditional echelle
spectrograph techniques operate by spreading a single stellar spectrum
over an entire CCD detector in multiple orders, here the spectrum only
takes up one strip along the detector. Spectra from multiple stars can
therefore be lined up at once on a single detector. In
combination with a wide field multi-fiber telescope, this makes
multi-object RV planet surveying possible
\citep{Ge2002,GeErskine02,SuvrathSPIE}. The multi-object Keck ET
instrument based on the DFDI technique is one of the first instruments to
be built with this purpose
\citep{GeMARVELSWhitePaper}.\footnote{Comparable traditional
  dispersive multi-object instruments are the VLT GIRAFFE and
  UVES/FLAMES spectrographs \citep{VLTFLAMES},
  and the MMT Hectochelle \citep{Hectochelle}}

The very high levels of precision required for planet detection and
the difficulty of directly measuring absolute wavelengths mean that
some kind of stationary reference spectrum is invariably used as a
calibration. Various types of fiducial reference have been employed to
overcome these problems \citep[e.g.][]{Griffin&Griffin,HFcell}, but the
references of choice have generally become ThAr emission lamps
\citep[]{elodie} and iodine vapor absorption cells placed within the
optical beam path \citep{Butler3ms}. In this respect, the ET
instruments are the same, and we discuss the use of such references
with the DFDI technique in this paper.

\subsection{A Brief History}

The idea of using the combination of a Michelson interferometer with a
postdisperser was first proposed for precision Doppler planet searches
by D.~J.~Erskine in 1997, at Lawrence Livermore National Laboratory
\citep{LLNL,Ge2002,GeErskine02,Erskine03}. The same approach is being
followed by \citet{TEDI} in the infra-red, in an attempt to find
planets around late-type stars. A similar approach is discussed by
\citet{Mosser03} for asteroseismology and the measurement of stellar
oscillations; more recently the technique has also been adopted for
the USNO Dispersed Fourier Transform Spectrograph (dFTS) instrument
\citep{USNOdFTS,USNOdFTS2} (in this last case, the interferometer delay is also
varied so that high resolution spectra can be reconstructed in
addition to extracting Doppler shift information -- see also
\citet{Erskine04ResBoosting}).

The idea of dispersed interferometry itself is by no means new:
Michelson himself recognized the use of interferometers for
spectroscopy \citep{MichelsonLightWaves}, and even proposed combining
a disperser in series with a Michelson interferometer. In this case
the disperser, a prism, was placed {\em before} the interferometer,
allowing only a narrow bandwidth of light to enter the interferometer
in the first place. In what was likely the first realization of a
DFDI, \citet{EdserButler} placed a Fabry--Perot type interferometer in
front of a spectrograph\footnote{It was mistakenly stated in
  \citet{51peg} that \citet{EdserButler} used a Michelson rather than
  a Fabry--Perot interferometer, which has certain disadvantages in
  this application \citetext{D.~J.~Erskine 2005, private communication}.}
to produce dispersed fringes (effectively an interferometer comb - see
section \ref{sec:comb}), which they used as a fiducial reference for
measuring the wavelengths of spectral lines. Such dispersed fringes
were later to become known as `Edser--Butler fringes' \citep{Lawson}.

Somewhat later, along with the development of P.~Connes' SISAM
\citep[``spectrom\`{e}tre interf\'{e}rentiel \`{a} s\'{e}lection par
l'amplitude de modulation,'' described in][]{Jacquinot}, various
combinations of interferometers with dispersers began to be seen in
the field of astronomy. Examples include \citet{Geake59}, using a
Fabry-Perot in front of a spectrograph to increase throughput; and the
later SHS \citep[``spatial heterodyne spectroscopy,''][] {Harlander92}
and HHS \citep[``heterodyned holographic
spectroscopy,''][]{Frandsen93, Douglas97} techniques, using {\em
  internally} dispersed interferometers, where the interferometer
mirrors were replaced with gratings. \citet{Barker&Hollenbach}
outlined an early example of the use of true fixed-delay
interferometry for velocimetry, measuring the velocities of
laser-illuminated projectiles in the laboratory. The use of a
Michelson interferometer for actual astronomical RV measurements was
proposed shortly afterward by \citet{Gorskii&Lebedev} and
\citet{Beckers&Brown}. \citet{Forrest&Ring} also proposed using a
Michelson interferometer with a fixed delay for high-precision Doppler
measurements of single spectral lines for the detection of stellar
oscillations, and more recent examples of similar spectroscopic
techniques include \citet{Connes85} and \citet{McMillan93,
  McMillan94}. Others have also used similar techniques for Doppler
{\em imaging} over very narrow bandpasses, notably the WAMDII
(wide-angle Doppler imaging interferometer) and GONG (Global
Oscillation Network Group) projects \citep{Shepherd85, GONG}.

Many of these interferometric instruments, however, suffered from the
limitation of having an extremely narrow bandpass, tending to limit
their application to only bright targets. The DFDI technique used in
the ET instruments allows for an arbitrarily wide bandpass, limited
only by the spectrograph capabilities, while still retaining the high
resolution spectral information needed for precision velocity
measurements. The first such DFDI instruments were built at the
Lawrence Livermore National Laboratory and the Lick 1m telescope
between 1997 and 1999, and were reported in \citet{LLNL} and
\citet{GeErskine02}. The ET project was undertaken shortly after.

\subsection{The ET Project}
The ET project began at Penn State University in 2000, continuing
at the University of Florida from 2004. Early lab tests were
performed at Penn State, and prototype test runs were conducted at the
McDonald Observatory Hobby-Eberly Telescope in late 2001, and at the
Palomar 200 inch telescope in early 2002
\citep{Ge03SPIE4838,suvrathsthesis}.

Two ET instruments have now been built: the single-object prototype ET
\citep{SPIE2004,monolithic2008}, permanently installed at the KPNO
2.1m telescope in 2003 after a temporary test run in August 2002; and
the multi-object Keck ET, first installed at the APO Sloan 2.5m
telescope in March 2005, upgraded and moved to a more stable location
at the same telescope later that year, and then further upgraded and
fully installed as facility instrument housed in its own custom-built
room in September 2008. The latter instrument will function as the
workhorse for the SDSS III ``Multi-object APO Radial Velocity Exoplanet
Large-area Survey'' \citep[MARVELS;][]{GeMARVELSWhitePaper}.

Proof of concept was achieved using the KPNO ET with the first DFDI
planet detection, a confirmation of the known companion to \object{51
  Pegasi} \citep{51peg}. Our first planet discovery, \object{HD
  102195b} (ET-1), was also later made using this instrument
\citep{ET1}. The multi-object Keck ET is a full scale instrument
developed to satisfy the survey requirements laid out in section
\ref{theneed}, and it is anticipated that it will be able to make a
significant contribution to the field of extrasolar planet searches
over the next decade \citep{GeMARVELSWhitePaper}.

%% file: section2.tex
\section{Instrument Principles and Theory}\label{sec:CHtheory}
Although various forms of the DFDI have been employed before, the
concept, particularly in its specific application to exoplanet
finding, is rather new. Much of the work in understanding the data
from the instrument has therefore involved coming to a full
understanding of the physics of the instrument itself. Related theory
is discussed in a number of sources \citep[for
example][]{Goodman,LLNL,Lawson,Ge2002,GeErskine02,Erskine03,
  Mosser03,SPIE2003}; an attempt is made here to draw together, expand
on, and precisely state the theoretical material needed for a complete
understanding of the instrument, and to provide an overview of the
physics underlying the instrument's working from the perspective of
precision RV planet detection. The approach taken here allows for
some important insights, particularly regarding certain errors
arising from the use of a common-path fiducial reference spectrum such
as that from an iodine gas absorption cell. In addition we derive in
section \ref{sec:CHerrors} some useful general formulae that can be
applied to estimate analytically the magnitude of both these and a
number of other types of error generic to the technique.

Taken together, this discussion should provide some of the
fundamentals necessary for understanding and interpreting DFDI
data. Appendix \ref{sec:APanotherapproach} gives a derivation showing
the relation to the approach employed by \cite{Erskine03}, to which
the approach here is complementary.

\subsection{Formation of a Fringing Spectrum}\label{sec:DFDIdescription}

\begin{figure*}[tbp] \centering \epsscale{0.7}
\plotone{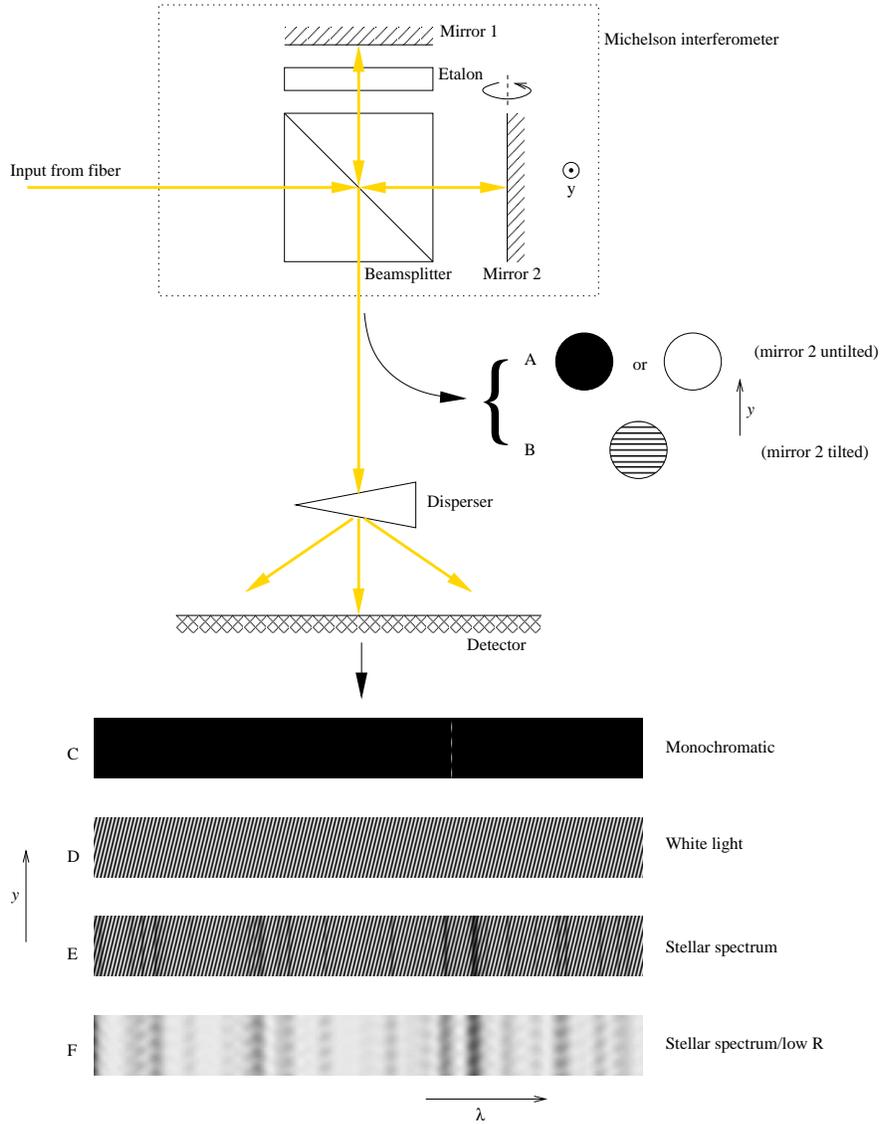}
\caption[Dispersed interferometer schematic]{\label{fig:ETschematic}Dispersed
  interferometer schematic. $y$ corresponds to position in the slit
  direction (directed out of the page in the interferometer
  schematic), and $\lambda$ indicates wavelength in the dispersion
  direction. A) Output from interferometer alone with monochromatic
  light input, and mirror 2 untilted. B) The same with mirror 2 tilted
  along the axis in the plane of the page, as shown. C) Image on
  detector with monochromatic light at very high resolution. One
  fringed emission line is seen. D) Detector image with white light
  input. E) Image with stellar spectrum input. F) As for E but at low
  resolution.}
\end{figure*}

Figure \ref{fig:ETschematic} shows a highly simplified schematic of a
DFDI, consisting of the two main components, a fiber-fed Michelson
interferometer and a disperser, followed by a detector. Light input
from the fiber is split into two paths along the arms of the
interferometer and then recombined at the beamsplitter. The output is
fed to the disperser, represented for convenience as a prism, though
generally this will be a spectrograph. An etalon is placed in one of
the interferometer arms to create a fixed optical path difference (or
`delay'), $d=d_0$, between the two arms, while allowing for adequate
field widening \citep{fieldwidening,monolithic2008}. $d_0$ is
typically on the order of millimeters. In practice, an iodine vapor cell
can also be placed in the optical path before or after the
interferometer to act as a fiducial reference (section
\ref{sec:I2separation}).

Inputting a wide collimated beam of monochromatic light into the
instrument with both interferometer mirrors exactly perpendicular to
the light travel path will give either a bright or a dark fringe at
the output of the interferometer (figure \ref{fig:ETschematic}A),
depending on whether the exact path difference $d$ between the two
arms corresponds to constructive or destructive interference. If we
were to scan one of the mirrors back and forth, the flux at the
interferometer output would vary sinusoidally as a function of $d$. If
we now tilt this mirror along the axis in the plane of the page, we
effectively scan a small range of delays along the $y$ direction
(i.e. perpendicular to the axis of the tilt and in the plane of the
mirror, corresponding to the slit direction in the
spectrograph). Hence we would see a series of parallel bright and dark
fringes, now varying sinusoidally as a function of
$y$.\footnote{Another way of sampling the fringes is to scan the
  interferometer delay in very small steps \citep[see][]{Erskine03}:
  this allows for certain advantages in calibration as well as a
  one-dimensional spectrum which requires less detector real-estate,
  but comes at the disadvantage of requiring an actively controlled
  interferometer. The principles are the same, however.}

Consider first a very high (actually infinite) resolution spectrograph
disperser for the sake of argument: following the beam through until
it reaches the
detector plane would result in a single emission line with fringes
along the slit direction, as shown in figure
\ref{fig:ETschematic}C. Switching the input spectrum to white light,
which can be thought of as a continuum of neighboring delta functions
in wavelength ($\lambda$) space, leads to a similar fringe pattern on
the detector at every wavelength channel. Due to the fact that, in
terms of number of wavelengths, the optical path difference is
different for different wavelengths, each fringe is slightly offset in
phase from its neighbors (and very slightly different in
period). This gives rise to the series of parallel lines known as the
interferometer `comb,' shown in figure \ref{fig:ETschematic}D. Going
further and inputting a stellar spectrum into the instrument would
simply give the product of the stellar spectrum and the comb, as in
figure \ref{fig:ETschematic}E. Finally, changing to the real case of a
low or medium resolution spectrograph as for an ET-type instrument,
the comb is no longer (or barely) resolved, and we see a spectrum like
that in figure \ref{fig:ETschematic}F. Such a spectrum is sometimes
referred to as a spectrum ``channeled with fringes,'' also known as
Edser-Butler fringes \citep{EdserButler,Lawson,Ge2002}. The remaining
fringes contain high spatial frequency Doppler information that has
been heterodyned down to lower spatial frequencies by the
interferometer comb \citep{Erskine03,suvrathsthesis}. It is this
heterodyning that allows for the use of a low-resolution spectrograph
at low dispersion, and is the key to the DFDI technique.


\subsection{Fringe Phase and Visibility}\label{sec:phaseandvis} Above
we outlined a simple intuitive way of understanding the formation of
the DFDI fringing spectrum. For a full mathematical description, we
proceed by a slightly different route. Each wavelength channel on the
detector has an associated sinusoidal fringe running along the slit
direction, where by `channel,' we mean specifically an infinitesimally
wide strip of the spectrum along the slit direction at pixel position
$j$, where $j$ need not necessarily be an integer.  A given fringe has
an associated phase and visibility, where visibility is a measure of
the contrast in the fringe, defined as the ratio of the amplitude of
the fringe to its central (mean) flux value. Equivalently, this can be
stated as
$(I_{\mathrm{max}}-I_{\mathrm{min}})/(I_\mathrm{max}+I_\mathrm{min})$,
where $I_\mathrm{max}$ and $I_\mathrm{min}$ are the maximum and
minimum flux values in the fringe \citep{MichelsonLightWaves}. Here we
introduce the concept of a `whirl' \citep{LLNL}: the phase and
visibility for a fringe can together be thought of as representing a
vector, with the visibility representing the magnitude. These
quantities can be determined in a number of ways; in general we simply
fit a sinusoid. An ensemble of such vectors representing a full
spectrum of channels is called a whirl. The whirl is the directly
measured quantity from a fringing spectrum and contains the
information relevant to velocity determination. Vector operations such
as addition, subtraction, and scalar products can be performed on
these whirls just as for the individual vectors \citep{LLNL}.

To understand what determines the values of the phase and visibility
for a fringe, we can consider the contribution from each wavelength of
light to a particular channel on the detector (remembering that
although the channel is infinitesimally wide in its spacial extent in
the dispersion direction, it still has a finite bandwidth). Each
contributing wavelength has passed through the interferometer, and for
an ideal interferometer, will contribute a sinusoid of 100\%
visibility like that in figure \ref{fig:ETschematic}C. The flux of
these sinusoids on the detector can each be described by
$\Re\{1+\exp(\mathrm{i} 2\pi d/\lambda)\}$, where $d$ varies linearly
with position $y$ along the length of the slit, and $\Re\{\ldots\}$
represents the real part of a complex expression. Since the
spectrograph has finite resolution, a narrow band of such wavelengths
will contribute to any given channel, owing to the overlap of line
spread functions (LSF's) from neighboring wavelengths. The measured
fringe along the slit direction is a continuous summation of those
sinusoids, weighted by the flux of the spectrum contributing to that
channel, $Q_j(\lambda)$, where $Q$ is given by the product of the
power spectrum coming into the instrument and the spectrograph
response function at that channel on the detector. We use the term
`spectrograph response function' throughout to refer to the light
throughput as a function of wavelength at a given infinitesimal point
on the detector, or equivalently, at a given channel in the image on
the detector. (This is distinct from, though closely related to, the
LSF -- see appendix \ref{sec:AppLSF}.)

Switching from wavelength to wavenumber $\kappa \equiv 1/\lambda$, and
dropping the $j$ subscript for simplicity, the summation of sinusoids
can be expressed as:
\begin{equation}\label{eqn:howitallbegins} I(d)\quad = \quad \int
Q(\kappa)\,\Re\{1+ \mathrm{e}^{\ii 2\pi \kappa d}\}\,\dd \kappa \quad
= \quad \int{Q(\kappa)\,\dd \kappa} + \Re\left\{\int
Q(\kappa)\mathrm{e}^{\ii 2\pi \kappa d}\,\dd\kappa\right\},
\end{equation} where $I(d)$ is the measured flux along the slit
direction. The first term on the right hand side is simply the total
integrated flux in the channel, which must be real valued. The second
term can immediately be identified as the real part of a Fourier
transform, $\Re\{\FT[Q]_d\}$, with delay as the conjugate variable to
wavenumber, and shows the close relationship between DFDI instruments
and Fourier transform spectroscopy \citep{Jacquinot}.

Normalizing by dividing through by the total flux, we can define the
complex quantity $\boldsymbol{\alpha}$ such that
\begin{equation}\label{eqn:normalisedfringe} I_\mathrm{norm}(d) = 1 +
\Re\left\{\frac{\FT[Q(\kappa)]_d}{\int{Q(\kappa)\,\dd\kappa}}\right\}
= 1+\Re\{\boldsymbol{\alpha}\},
\end{equation} where
\begin{equation}\label{eqn:complexvis} \boxed{\boldsymbol{\alpha}
\equiv \alpha\ee^{\ii\phi_\alpha}
\equiv\frac{\FT[Q(\kappa)]_d}{\int{Q(\kappa)\,\dd\kappa}}.}
\end{equation} This is the fundamental equation for DFDI fringe
formation: the quantity $\boldsymbol{\alpha}$ is the `complex degree
of coherence' \citep{Goodman}, and describes the phase, $\phi_\alpha$,
and amplitude, $\alpha$, of the normalized fringes (i.e. the
visibility), as a function of $d$ and the input
spectrum. $\boldsymbol{\alpha}$ is referred to here as the
{\em complex visibility.}\footnote{The quantity is generally represented by the
  letter $\boldsymbol{\gamma}$ in the literature cited. We use
  $\boldsymbol{\alpha}$ here instead purely for clearer distinction
  between bold-faced vector and regular-faced amplitude
  representations.}  More rigorous derivations of this can be
found in \citet[ch. 5]{Goodman} and \citet{Lawson}, but this
explanation is adequate for our purposes.

\begin{figure*}[htbp] \epsscale{1.0} 
    \plotone{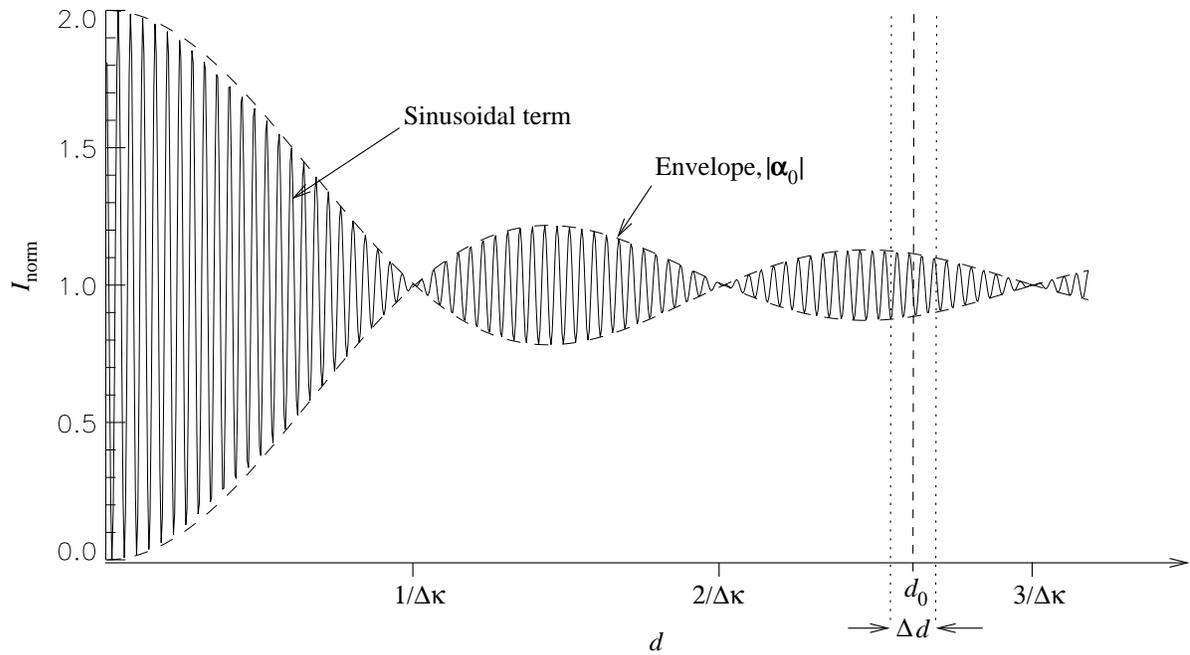}
    \caption[Interferogram showing the coherence envelope due to a
    rectangular band pass]{\label{fig:envelope}Interferogram showing
    the coherence
    envelope due to a rectangular band pass modulating the
    sinusoidal fringe. Along the slit direction of a fringing
    spectrum, a very small part of the interferogram is sampled over
    the range $d_0\pm\Delta d/2$.}
\end{figure*}

In order to understand the actual form of the fringes seen in a DFDI,
it is important to realize that the portion of spectrum contributing
to any given channel, $Q$, has a very narrow passband (for the ET
instruments, $\Delta \lambda /\lambda \sim 1\ang /5000\ang = 2\times
10^{-4}$). We imagine $Q$ as being a shifted version of a function
$Q_0$, where $Q_0$ has characteristic width $\Delta \kappa$ and is
centered at zero wavenumber. We shift $Q_0$ in wavenumber so that its
center falls at wavenumber $\kappa = \overline{\kappa}$, and
we have $Q(\kappa) = Q_0(\kappa - \overline{\kappa})$. By the Fourier shift
theorem we can write:
\begin{equation}\label{eqn:modulatedsin}
\FT[{Q}]_d = \FT[Q_0(\kappa - \overline{\kappa})]_d = \ee^{-\ii2\pi d\overline{\kappa}}\FT[{Q_0}]_d.
\end{equation}
The right hand side shows two components. The exponential term
represents a linear phase variation with delay, varying on the scale
of the period $1/\overline{\kappa}$. The second term, the Fourier
transform, represents a modulation of this signal. By the reciprocal
scaling property of Fourier transforms, the second term can be
expected to vary on minimum length scales of the order of the
reciprocal of the width of $Q_0$, that is, on scales of $1/\Delta
\kappa$. Since $1/\Delta \kappa \gg 1/\overline{\kappa}$, equation
\ref{eqn:modulatedsin} represents a sinusoidal fringe of frequency
$\overline{\kappa}$ modulated by a slow variation in both phase and
amplitude. To see this more clearly, we can substitute equation
\ref{eqn:modulatedsin} into the first expression on the right hand
side of equation \ref{eqn:normalisedfringe} and
write:
\begin{equation}\label{eqn:normalisedfringe2}
I_\mathrm{norm}(d)=1 +
\frac{\Re\left\{ \ee^{-\ii2\pi d\overline{\kappa}}\FT[{Q_0}]_d\right\}}{\int{Q(\kappa)\,\dd\kappa}}.
\end{equation}
If we define
\begin{equation}\label{eqn:envelope}
\boldsymbol{\alpha}_{0}(d) \equiv \alpha_{0}(d) \ee^{\ii\phi_{\alpha_{0}}(d)} \equiv
\frac{\FT[Q_0]_d}{\int{Q(\kappa)\,\dd\kappa}},
\end{equation}
we can rewrite equation \ref{eqn:normalisedfringe2} as:
\begin{eqnarray}
I_\mathrm{norm}(d) & = & 1+\Re\left\{\boldsymbol{\alpha}_{0}(d)\ee^{-\ii2\pi
                   d\overline{\kappa}}\right\} \nonumber
\\
                   & = & 1 + \alpha_{0}(d) \cos(2\pi d\overline{\kappa} -
                   \phi_{\alpha_{0}}(d))\label{eqn:normalisedfringe3}
\end{eqnarray}
(where we have simplified the negative in the cosine term using the
symmetry of the cosine function). This clearly shows the form of the
fringe. Over large ranges of $d$, the fringe appears like a `carrier
wave,' given by the cosine term, that is slowly modulated in phase and
amplitude by an envelope $\boldsymbol{\alpha}_{0}$ \citep[the
`coherence envelope',][]{Lawson}. Over the length of the slit
direction on the detector, we sample only a very small range of
delays, $d_0-\Delta d/2 \: \le \: d \: \le \: d_0+\Delta d/2$, where
$d_0$ is determined by the interferometer etalon, as before, and
$\Delta d$ is typically a few wavelengths. Over this range, the
variation in $\boldsymbol{\alpha}_{0}$ is small as we show below, so
we see only an approximately uniform sinusoid (see figure
\ref{fig:envelope}) along a single wavelength channel on the
detector. In measuring the phase and visibility of this fringe, we
essentially make a measurement of $\boldsymbol{\alpha}_{0}$ at the
fixed delay $d=d_0$. The phase offset of the sinusoid is determined by
the argument of $\boldsymbol{\alpha}_{0}$, $\phi_{\alpha_{0}}$. The
measured (absolute) fringe visibility is simply the amplitude of the
normalized fringe, $\alpha_{0}$.

In general, we can estimate a rough order of magnitude for the
fractional change in the magnitude of the visibility between
consecutive sinusoid peaks by comparing the variation length scales:
to order of magnitude, we can expect $\alpha_{0}$ to vary by of order
$\alpha_{0}$ on scales of $1/\Delta \kappa$, so that over one period
of the sinusoid, $1/\overline{\kappa}$, it will vary by $\Delta
\alpha_{0} = \alpha_{0}\Delta \kappa/\overline{\kappa}=\alpha_{0}/R$,
where $R$ is the spectrograph resolution. Since for any input
spectrum, $1/\Delta\kappa$ determines the fastest variation scale for
the envelope, this represents an upper limit. For the ET instruments,
$R\sim 5000$, so that over the length of the slit (a few fringes)
$\Delta \alpha_{0}/\alpha_{0}\sim 10^{-3}$. In practice such a small
variation will usually be significantly below the measurement errors
in fringe phase and visibility due to photon noise for even the
brightest sources, and would correspond to a final velocity error of
$\sim 0.1\ms$ for an instrument similar to the KPNO
ET.\footnote{Assuming $\sim 1000$ independent channels, phase-velocity
  scaling factor $\Gamma\approx3300\msrad$ (see section
  \ref{sec:phasetovelocity}), and using the relationship between phase
  error and visibility error shown in section \ref{sec:photonerrors},
  equation \ref{eqn:phaseerrvsviserr}, so that the expected error is
  $\Gamma \varepsilon_{\phi,j}/\sqrt{1000} =
  10^{-3}\Gamma/\sqrt{1000}$.}  
Even in the event that it is desired
to reach such extremely high signal-to-noise ratios (S/Ns), it is in
principle a simple matter to fit extra parameters to allow for
non-uniformity of the sinusoidal fringe, although this has not been
attempted with the ET instruments.

In figure \ref{fig:envelope}, the varying amplitude of the modulating
coherence envelope, $\boldsymbol{\alpha}_{0}$, is illustrated explicitly, and we see
how measuring the fringe over a narrow range of delays $\Delta d$
around $d_0$ gives an approximately uniform sinusoid. This corresponds
directly to the image seen along the length of the slit direction in a
given channel on the detector. For illustration the very simple case
is shown of white light with through a rectangular bandpass with no
absorption lines, so that $Q$ (and therefore $Q_0$) is a top-hat
function. $\boldsymbol{\alpha}_{0}(d)$, therefore, is the corresponding Fourier
transform, a sinc function, with zeros at $d=n/\Delta\kappa$
$(n\in\mathbb{Z}^+$), which modulates a sinusoid of period
$1/\overline{\kappa}$. In practice the passband, $\Delta
\kappa/\overline{\kappa}$, will be very narrow, so that the variation
of $\boldsymbol{\alpha}_{0}$ will be much slower compared to the sinusoid than
suggested in the figure, and the sinusoid itself will be highly
uniform over $\Delta d$ (i.e. over the length of the slit).

For a more complicated input spectrum, such as that from a star with
its multitude of absorption lines, and for a more realistic LSF, the
coherence envelope will generally also have a much more complicated
shape, though the variations will still be slow in $d$ and therefore
close to uniform along the slit (i.e. within the upper limit discussed
above, since the width of the resolution element still determines the
fastest variation scale). Each channel will have its own
unique piece of spectrum contributing to it, and therefore each will
have its own particular phase and visibility. It is this that gives
rise to the varied patterns of fringes that are seen in the final
fringing stellar spectra (e.g. figure \ref{fig:ETschematic}F).

In practice the profile in the slit direction will also be modulated
in amplitude by a slit illumination function, but this can be
calibrated out or modeled during the fringe fitting, and has no
effect on fringe visibility. Though this can present its own
practical challenges for data reduction, the illumination function is
neglected here for simplicity, and taken to be uniform and equal to
unity.

As an aside we note that $\boldsymbol{\alpha}_{0}$ and $\boldsymbol{\alpha}$
are very closely related: from their respective definitions in
equations \ref{eqn:envelope} and \ref{eqn:complexvis}, $\boldsymbol{\alpha}_{0}(d)
= \ee^{\ii 2\pi d\overline{\kappa}}\boldsymbol{\alpha}(d)$. The only
difference is a phase offset, which, for a given channel $j$ at
wavenumber $\overline{\kappa}_j$ and fixed delay $d=d_0$, is constant
-- that is to say, $\alpha_{0}=\alpha $ and $\phi_{\alpha_{0}} = \phi_\alpha + 2\pi
d_0\overline{\kappa}_j$. Since the instrument is to be used purely for
differential measurements, the zero point from which phases are
measured is somewhat arbitrary and has no physical significance: we
are concerned with {\em changes} in phase over time, which will affect
both $\boldsymbol{\alpha}_{0}$ and $\boldsymbol{\alpha}$ in the same way. For the
analyses presented hereafter, the difference between $\boldsymbol{\alpha}_{0}$ and
$\boldsymbol{\alpha}$ is therefore not of great significance, and
either can equally well be thought of as the complex
visibility. However, for the sake of consistency,
$\boldsymbol{\alpha}$ is generally intended by the term.

\subsection{From Phase to Velocity} \label{sec:phasetovelocity}
To recap, in general, for a given channel $j$ on the detector, the
complex visibility of the measured fringe is given as in equation
\ref{eqn:complexvis} \citep[or see][chap. 5]{Goodman}. We can rewrite
this as:
\begin{equation}\label{eqn:cplxvis}
\boxed{
  \boldsymbol{\alpha} = \frac{\mathcal{F}[P_\kappa w_{\kappa j}]_{d=d_0}}{\mathcal{F}[P_\kappa w_{\kappa j}]_{d=0}}
  = \frac{\mathcal{F}[P_\nu w_{\nu j}]_{\tau=\tau_0}}{\mathcal{F}[P_\nu w_{\nu j}]_{\tau=0}},
}
\end{equation}
where $\boldsymbol{\alpha}$ is the complex visibility (or complex
degree of coherence), a vector quantity whose phase represents the
phase of the measured fringe, and whose magnitude (from 0 to 1)
represents the absolute visibility of the measured fringe;
$\mathcal{F}[\ldots]_{\ldots}$ represents a Fourier transform
evaluated at interferometer path difference $d$, or time delay $\tau$,
where $d=c\tau$ and $c$ is the speed of light; $P$ is the input
spectrum; and $w_j$ is the response function for that particular
channel on the detector, so that the spectrum contributing to the
channel is given by $Q_j = Pw_j$ as before. We take $d$ to be fixed at
a value $d_0$ (for the purposes of the calculations here, the small
difference in $d$ across the length of a sinusoidal fringe is of no
consequence). Subscripts are added to explicitly indicate functions of
wavenumber, $\kappa$, or optical frequency, $\nu = c\kappa$: we note
that the equation is completely equivalent in $\kappa$ space with $d$
as the conjugate variable, or in $\nu$ space with $\tau$ as the
conjugate variable. In general the form being used will be implicit
from the context, so we drop these subscripts. We have also replaced
the integral over the flux in the denominator with the Fourier
transform at zero delay, which is mathematically equivalent (this
fact is made use of a number of times later on in this
analysis). All the necessary mathematics for determining Doppler
shifts and for dealing with the combination of the star and fiducial
reference spectra (see section \ref{sec:I2separation}) derive from
this formula.

The key to the DFDI RV technique is the fact that Doppler
shifts of the spectrum result in directly proportionate phase shifts
of the fringes. This is a direct consequence of the Fourier shift
theorem \citep[see e.g.][]{Erskine03}. If the spectrum shifts such
that $P(\kappa) \rightarrow P^{\prime}(\kappa) \equiv
P(\kappa+\Delta\kappa)$, and we correctly follow the shift in the
dispersion direction (so that we now compare to the wavelength channel
corresponding to $w_{j+\Delta j} = w_j(\kappa+\Delta\kappa)$ --
assuming that the spectrograph response function maintains the same
form in nearby channels, and noting that $\Delta j$ is not necessarily
an integer), then the shift theorem gives:
\begin{equation}
\boldsymbol{\alpha}^{\prime} =
\frac{\mathcal{F}[P(\kappa+\Delta\kappa)w_j(\kappa+\Delta\kappa)]_{d=d_0}}{\mathcal{F}[P(\kappa+\Delta\kappa)w_j(\kappa+\Delta\kappa)]_{d=0}}
=
e^{i2\pi\Delta\kappa d_0}\frac{\mathcal{F}[P(\kappa)w_j(\kappa)]_{d=d_0}}{\mathcal{F}[P(\kappa)w_j(\kappa)]_{d=0}}
= e^{i2\pi\Delta\kappa d_0}\boldsymbol{\alpha}.
\end{equation}
In other words, we have a phase shift of
$\Delta\phi = 2\pi d_0\Delta\kappa$. By comparing the measured phase of the
new fringes $\boldsymbol{\alpha}^{\prime}$ with the previously unshifted
ones, $\boldsymbol{\alpha}$, it is thus possible, in this simple case
where there is no superposed reference spectrum and the instrument is
perfectly stable, to derive the Doppler
shift {\em without any explicit knowledge of the underlying high resolution
spectrum, or of the spectrograph LSF.} Using the Doppler shift equation
$\Delta\kappa/\kappa \approx -\Delta v/c$, where $v$ represents velocity,
conventionally positive in the direction away from the observer, we can write:
\begin{equation}\label{eqn:p2v}
\Delta\phi = 2\pi d_0\Delta\kappa = -\frac{2\pi d_0\kappa\Delta v}{c} =
-\frac{2\pi d_0}{c\lambda} \Delta v \equiv \frac{\Delta v}{\Gamma},
\end{equation} 
where, $\Gamma$, the phase-velocity scaling factor which gives the
proportionality between phase shift and velocity shift, is defined as:
\begin{equation}\label{eqn:Gamma}
\boxed{\Gamma \equiv -\frac{c\lambda}{2\pi d_0}.}
\end{equation}
By combining the many measurements of the phase shift $\Delta \phi$
from each channel, $j$, (allowing, if necessary, for the wavelength
dependence of $\Gamma$), a very high precision measurement of the
differential Doppler velocity shift, $\Delta v$, can be made.


\subsection{The Interferometer Comb} \label{sec:comb}

The interferometer comb, mentioned in figure \ref{fig:ETschematic} and
the corresponding text, is really just a special case of the
discussion in section \ref{sec:phaseandvis}, where the input spectrum
to the instrument is purely white light continuum. In that case $Q$,
the product of the input spectrum and the spectrograph response
function, is itself equal to the spectrograph response function. The
comb is therefore purely a consequence of the response function,
arising naturally from equation \ref{eqn:complexvis}. In fact, the
example used of the top-hat function for $Q$ is a reasonable first
approximation for the LSF, and so also for the response function (see
appendix \ref{sec:AppLSF}), for a spectrograph where the slit width
dominates the resolution. The interferogram in figure
\ref{fig:envelope} is thus a reasonable representation of the
behavior of the interferometer comb at finite resolution.

We can see from this that by appropriately choosing the delay and
spectrograph slit width we can null out the interferometer comb by
finding a minimum in the envelope. Early experiments changing the slit
width and delay with ET prototypes did indeed show this kind of
sinc-like variation in the comb visibility. This becomes important
when using a superimposed reference spectrum, as in section
\ref{sec:additionapprox}.

It is also instructive to consider an idealized infinite resolution
spectrograph. In this case, the response function, $w_j$, becomes
a delta function, so that $Q_j$ is also a delta function for all
channels $j$. By equation \ref{eqn:envelope}, given that $Q_0$ is the
delta function shifted to $d=0$, the coherence envelope,
$\boldsymbol{\alpha}_{0}(d)$, is the normalized Fourier transform of this delta
function: $\boldsymbol{\alpha}_{0}(d)=1$ at all delays. Equation
\ref{eqn:normalisedfringe3} then gives the very simple form of the
resulting interferogram:
 \begin{equation}\label{eqn:comb}
   I_\mathrm{norm} = 1+\cos(2\pi d\kappa) 
 \end{equation}
 where we have stopped representing $\overline{\kappa}$ as a mean value
 since the width of the channel is negligible.

 This 100\% visibility `infinite resolution' comb is the underlying
 form for any DFDI comb. Lowering the resolution will reduce the
 visibility from 100\% at the given fixed delay, as in the example of
 figure \ref{fig:envelope}, with perhaps an overall phase offset
 depending on the symmetry of the spectrograph response function (and
 uniform to the extent that the response function and LSF are uniform
 across all channels).

 The infinite-resolution comb can also be thought of as a an
 interferometer transmission function. In introducing the instrument
 (section \ref{sec:DFDIdescription}), we first described the formation
 of the DFDI spectrum as a multiplication of the stellar spectrum and
 the infinite-resolution interferometer comb (i.e. interferometer
 transmission function), convolved with the LSF down to the
 spectrograph resolution. This is the approach adopted by
 \citet{Erskine03} and \citet{suvrathsthesis}, and both views are
 entirely equivalent. Following the Fourier transform approach
 outlined here, however, we can proceed somewhat further, and obtain
 some important insights in understanding systematic errors from the
 use of a simultaneous fiducial reference spectrum (section
 \ref{sec:I2separation}). In principle the Fourier transform approach
 can also be used to create simulated DFDI spectra without having to
 assume a uniform LSF at all wavelengths, which is difficult to do in
 the alternative approach. A derivation relating the two methods is
 outlined in appendix \ref{sec:APanotherapproach}.

\begin{figure}[tbp]
     \plotone{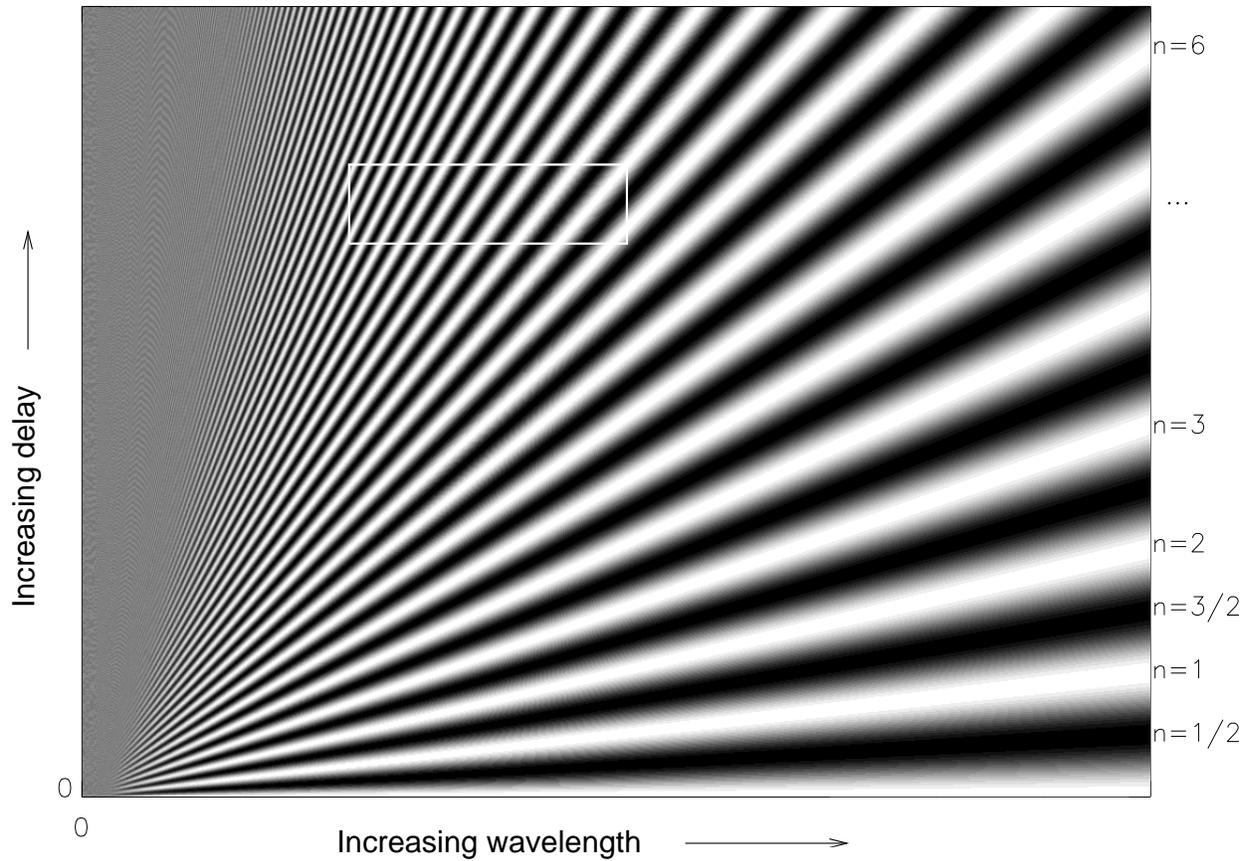}
     \caption[Simulated interferometer comb]{\label{fig:comb}Simulated
       interferometer comb, as a function of wavelength (corresponding
       with dispersion direction on detector) and delay (corresponding
       with slit direction). Setting a large interferometer delay and
       choosing the wavelength range over which the spectrum is
       observed selects a `window' in the comb (shown schematically)
       where the fringes are approximately parallel. The orders of some
       of the fringes, $n$, are shown down the right-hand side. In
       practice, the `window' chosen is at much longer wavelength and
       much higher order.}
\end{figure}

To show visually how the comb forms, it is depicted schematically in
figure \ref{fig:comb}, plotting contours of flux from equation
\ref{eqn:comb} as a function of wavelength $\lambda=1/\kappa$ and
delay $d$. Since $\lambda$ maps linearly to $x$ position on the
detector and delay maps linearly to $y$ position along the slit (at
least for an ideal spectrograph and interferometer), this also
represents the image that would be seen on the detector if the full
ranges could be sampled down to zero wavelength and zero delay. The
box in the figure schematically represents the segment of the
interferogram that we actually observe with the instrument: a series
of tilted parallel fringes (as shown in figure \ref{fig:ETschematic}D), with
a very slow wavelength dependency. For clarity, the figure is not to
scale: in practice, the delay is fixed to a much larger value so that
the fringes are observed at much higher order, $n$, and the
wavelengths observed are much longer, so that any real observed comb
is much denser and more uniform, and the variations with wavelength
much smaller.

\subsection{Calculating the Interferometer
  Delay} \label{sec:delaycalibration}

The interferometer delay, $d_0$, is determined by the etalon in the
interferometer, and must be known precisely in order to be able
to accurately translate from phase measurements to velocity
measurements. The best precision that can be obtained in RV
measurements is a trade-off between maximizing the phase-velocity
scale $\Gamma$ (so that a large phase shift results from a small
change in velocity) and maximizing the visibility of the fringes
(since higher visibility means more accurate measurements of the
fringe phases). Since the visibility of the fringes is determined by
the match between $d_0$ and the typical spectral line widths to be
observed, an optimal value of $d_0$ can be chosen to give the best
precision for the expected typical targets for the survey
\citep{Ge2002}. This is set at design time, and remains fixed for the
instrument.

Annual variations in the RV of a star can be as large as $\sim60\kms$ even
for an RV-stable target, owing simply to the orbital motion of the Earth
around the Sun (which dominates significantly over the Earth's
rotation). If we are to consider approaching precisions on the order
of $1\ms$ we therefore need to know $\Gamma$ to better than one part
in 60,000. Since $\Gamma$ depends directly on the interferometer delay
(equation \ref{eqn:Gamma}), determining $\Gamma$ is synonymous with
measuring the delay.

To a first approximation, the delay can be calculated from the
properties of the delay in the interferometer. For example, for a monolithic
interferometer with arm lengths $L_1$ and $L_2$ and refractive
indices $n_1$ and $n_2$ respectively, this is given by \citep{monolithic2008}:
\begin{equation}
d_0 = 2(n_1 L_1 - n_2 L_2),
\end{equation}

This depends on the assumption that there is negligible dispersion in
the etalon glass, i.e., that $n_1$ and $n_2$ are close to independent
of wavelength over the wavelength range of interest. Dispersion can in
fact be a significant effect, but the assumption should be good to a
few percent (\citealt{Barker1974}; D.~J.~Erskine 2001, private
communication), enough for an initial estimate. Accounting
fully for the dispersion and allowing $d_0$ to become a function of
wavelength, however, is essential where very high velocity precision
is required from large bandwidth observations.

A more precise measure of the delay can be determined simply by
counting fringes in the interferometer comb. We know from equation
\ref{eqn:comb} that the phase of the comb varies as $\phi = 2\pi
d\kappa = 2\pi d/\lambda$. Although this equation  is for a comb at infinite
resolution, the same variation will hold true at lower resolutions: a
spectrograph response function broader than a delta function will only
reduce the visibility of the interferogram, and possibly add an
overall phase offset to the entire interferogram (provided that the
shape of the response function is uniform across the
detector). Differentiating with respect to wavelength:
\begin{equation}
\frac{\partial\phi}{\partial\lambda} =
\frac{\partial (2\pi n)}{\partial\lambda} = -\frac{2\pi d}{\lambda^2},
\end{equation}
where $n=\phi/2\pi$ is the fringe order, giving:
\begin{equation}
d = -\lambda^2 \frac{\partial n}{\partial\lambda}.
\end{equation}
In other words, by counting the fringe density $\partial
n/\partial\lambda$ over wavelength, we can immediately calculate
$d_0$, and hence $\Gamma$. Since there is a $\lambda^{-2}$ dependence
in $\partial n/\partial\lambda$ itself, care needs to be taken to
account for the dependence properly when determining the fringe
density at a given wavelength. This may more easily be done in
wavenumber space instead, since the fringe density is uniform with
wavenumber, and $d=\partial n/\partial\kappa$.

In practice, counting fringes is often not easy, since the comb is
often barely resolved (usually by design). As long as the comb is not
under-sampled on the detector, this can be overcome by temporarily
using a narrower slit in the spectrograph, since in principle the
delay should only need to be determined once. Even so, it is usually
possible in practice only to count over a range of a few hundreds to
one or two thousand fringes. Counting along one row in the dispersion
direction of the comb therefore gives an accuracy on the order of one
part in 1000. Over a $60\kms$ variation, this is still only good to
the $60\ms$ level. Our method of choice in the past has been simply to
observe known stable reference stars over the time baseline of
interest and use their known apparent changes in velocity due to the
Earth's motion to calibrate $\Gamma$. Provided the reference stars are
genuinely stable, and they are positioned in the sky such that their
barycentric motions are large, this technique will provide an accuracy
in the determination of $\Gamma$ at least equal to the intrinsic RV
stability of the stars.

Other methods are under investigation which should allow more precise
measurement of the delay. By averaging fringe counts over many rows of
a wide spectrum, and further averaging over many frames, it may be
possible to achieve significantly sub-fringe counting accuracy
\citetext{J.~Wang et al., 2010, in preparation}. Other
techniques in development using a separate device to directly measure
the interferometer delay should provide a robust direct measurement
that obviates the need for more laborious empirical delay
determination \citetext{X.~Wan et al., 2010, in preparation}.


\subsection{Handling a Fiducial Reference
  Spectrum}\label{sec:I2separation}
\subsubsection{Multiplied Reference}\label{sec:additionapprox}

The extremely high sensitivity of the instrument means that numerous
instrumental effects can masquerade as velocity shifts. Tiny changes
in the interferometer delay due to thermal flexure, for example, will
appear as phase shifts in the fringe pattern. The image itself can
also shift as a whole on the detector in both the slit and the
dispersion directions. 

One way of accounting for these instrumental artifacts is to use a
fiducial spectrum from some known zero-velocity reference. The
simplest way to do this is to bracket the science data, either
spatially, running the fiducial spectra along a separate optical path
alongside the target spectrum; or temporally, alternating target
exposures and reference spectrum exposures along the same optical
path. Since the reference spectrum is stationary with respect to the
instrument, it will track instrument shifts, which can then be
subtracted from the measured stellar shift to reveal the star's
intrinsic motion. (Note that from equation \ref{eqn:p2v}, a change in
$d_0$ conveniently has mathematically exactly the same effect as a
change in velocity, $\Delta v$.) These approaches, however,
potentially suffer from errors due to their separation from the
science data: in the first case, because of non-common path errors due
to imperfect optics, and in the second case, because the fiducial
exposures are not tracking instrument drift contemporaneously with the
data.

An alternative approach is to insert an absorption reference into the
optical path -- in the case of the ET instruments in the past, a glass
cell filled with iodine vapor maintained at a fixed temperature, the
traditional reference of choice for RV planet searches. In this way
the reference spectrum is multiplied with the stellar spectrum. To do
this, for each target to be observed, two fringing `template' spectra
are taken, one being pure star with no reference in the beam path, and
the other pure reference (for ET, a pure iodine spectrum taken by
shining a tungsten continuum lamp through the cell).  These templates
are then used to separate out the stellar and reference components of
the combined star/reference data (referred to here as `data' or
`measurement' frames, as distinct from `template' frames). A formalism
is required to extract the reference and stellar spectra from the
combined spectrum. In order to proceed, we define the following
symbols:

\begin{itemize}

\item $j$ --- as before, the pixel number in the dispersion direction
  which identifies the column along which a fringe is measured in the
  slit
  direction, corresponding to a single channel. Strictly
  speaking, the channel is infinitesimally wide on the detector, so that
  $j$ need not necessarily be an integer. Since the spectrum is
  oversampled, however, it is often a reasonable simplification to
  think of the entire pixel column representing an infinitesimal
  sample in the dispersion direction (see appendix \ref{sec:AppLSF}).
\item $\mathbf{M}(j)$ --- the complex visibility vector (i.e. phase and
  absolute visibility) for a fringe at channel $j$ in a
  single Doppler measurement frame of combined star/reference data,
  an ensemble of such values for a spectrum across all $j$ comprising a `whirl.'
\item{$\mathbf{S}(j)$ --- the measured complex visibility for the star
 template at channel $j$.}
\item {$\mathbf{I}(j)$ --- the measured complex visibility for the reference
  template at channel $j$.}
\item {$\mathcal{M}(\lambda) \equiv C_\mathrm{m}(\lambda)M(\lambda)$
    --- the input spectrum for a combined star/reference data frame,
    where $C_\mathrm{m}$ represents a normalization, such as the
    continuum function, and $M$ is the normalized spectral
    density. $C_\mathrm{m}$ is assumed constant to a good
    approximation over the scale of the width of the response function
    $w$ (see below) and instrument LSF, and $0\le M \lesssim 1$.}
\item {$\mathcal{S}(\lambda) \equiv C_\mathrm{s}(\lambda) S(\lambda)$ --- the same for the star template spectrum.}
\item {$\mathcal{I}(\lambda) \equiv C_\mathrm{i}(\lambda) I(\lambda)$ --- the same for the reference template spectrum.}
\item {$s(\lambda), i(\lambda)$ --- such that  $S\equiv 1-s,\: I\equiv 1-i;\: 0\le (s,i) \le 1.$}
\item {$w(j,\lambda)$ --- the response function at position $j$ on the
  detector, i.e., the spectrum that contributes to an infinitesimally wide
  channel at the detector plane if perfect continuum light
  is passed through the instrument. (Note that $w$ is very closely related to
  the instrument LSF --- see appendix \ref{sec:AppLSF})}
\item $d$ --- the interferometer delay, fixed to a value of $d=d_0$,
  as usual.
\item $\Gamma$ --- phase/velocity scaling constant, also as before.
\item $\mathcal{F}[\ldots]_d$ --- as before, Fourier transform evaluated at
  interferometer path difference $d$.
\item $\widehat{\ldots}|_d$ --- shorter notation for Fourier transform, for convenience.
\item $[\ldots\otimes\ldots]|_d$ --- used to denote convolution,
  evaluated at a delay of $d$.
\end{itemize}

We assume for now the case where there is neither intrinsic Doppler
shift nor any instrument shift in either phase or in the dispersion
direction, for both star and reference components. We also assume no
photon shot noise. Here the aim is simply to reconstruct the data
whirl from the two template whirls. Once this is achieved, it is
conceptually a relatively trivial step to allow for shifted and noisy
data: the template whirls need only to be shifted iteratively in phase
and translated in the dispersion direction until a best-fit solution
is found, allowing the intrinsic stellar Doppler shift to be directly
calculated. This can be done using any standard least-squares method.

Following equation \ref{eqn:cplxvis}, the complex visibility measured
at detector channel $j$ for the two templates, $\mathbf{S}$ and
$\mathbf{I}$, and the combined star/reference data, $\mathbf{M}$, can
be written exactly as: \begin{equation}\label{eqn:star1} \mathbf{S} =
  \frac{\mathcal{F}[\mathcal{S}w]_{d_0}}{\mathcal{F}[\mathcal{S}w]_{0}}
  =
  \frac{[\widehat{\mathcal{S}}\otimes\widehat{w}]|_{d_0}}{[\widehat{\mathcal{S}}\otimes\widehat{w}]|_0},
\end{equation} \begin{equation}\label{eqn:iodine1} \mathbf{I} =
  \frac{\mathcal{F}[\mathcal{I}w]_{d_0}}{\mathcal{F}[\mathcal{I}w]_{0}}
  =
  \frac{[\widehat{\mathcal{I}}\otimes\widehat{w}]|_{d_0}}{[\widehat{\mathcal{I}}\otimes\widehat{w}]|_0},
\end{equation} \begin{equation}\label{eqn:data1} \mathbf{M} =
  \frac{\mathcal{F}[\mathcal{M}w]_{d_0}}{\mathcal{F}[\mathcal{M}w]_{0}}
  =
  \frac{\mathcal{F}[\mathcal{SI}w]_{d_0}}{\mathcal{F}[\mathcal{SI}w]_{0}}
  =
  \frac{[\widehat{\mathcal{S}}\otimes\widehat{\mathcal{I}}\otimes\widehat{w}]|_{d_o}}
  {[\widehat{\mathcal{S}}\otimes\widehat{\mathcal{I}}\otimes\widehat{w}]|_0}.
\end{equation}

The key lies in expressing equation \ref{eqn:data1} in terms of \ref{eqn:star1} and
\ref{eqn:iodine1}. This is made difficult by the convolutions, which
appear to require knowledge of the template spectra at all possible values of the
delay $d$ in order to be evaluated. The nature of the DFDI is such
that we measure it only at one value, $d_0$. An approximation
can be used to address this problem, which is described in
section \ref{sec:additionapprox}.


It is possible to rewrite the input spectrum as:
\begin{eqnarray}
  \mathcal{M} & = & A\mathcal{SI} \nonumber \\
              & = & A C_\mathrm{s} C_\mathrm{i}SI \equiv C^\prime SI  \nonumber \\
              & = & C^\prime(1-s)(1-i) \nonumber \\
              & = & C^\prime(1-s+1-i-1+si) \nonumber \\
              & = & C^\prime(S+I-1+si) \label{eqn:specexpanded},
\end{eqnarray}
where $A$ is a scaling constant to allow for difference in total flux
level between the templates and data, and $C^\prime \equiv
AC_\mathrm{s}C_\mathrm{i}$ is a constant over the width
of the response function. If we assume either $s$ or $i$ or both
$\ll 1$, then the `crosstalk' term, $si$, can be neglected. Since $i$
and $s$ essentially represent line depths, this means that we are
assuming either very shallow lines, or no significant overlap between
lines in the two different spectra. Keeping the crosstalk term in
place for now for completeness, however, we can continue, substituting equation
\ref{eqn:specexpanded} in the first expression on the right hand side
of equation \ref{eqn:data1}:
\begin{equation} \label{eqn:M1}
  \mathbf{M} =
  \frac{\mathcal{F}[\mathcal{M}w]_{d_0}}{\mathcal{F}[\mathcal{M}w]_{0}} =
  \frac{[\widehat{Sw} +
  \widehat{Iw}-\widehat{w}+\widehat{siw}]|_{d_0}}
  {[\widehat{Sw} +
  \widehat{Iw}-\widehat{w}+\widehat{siw}]|_0}.
\end{equation}
The factor $C^\prime$ has canceled because it is constant over the
width of the response function, and therefore can be taken outside the
Fourier transforms. The denominator of this equation represents a normalization, corresponding to
the total flux in channel $j$ on the detector. The term $\widehat{w}|_{d_0}$ in
the numerator is due to the interferometer comb, since if white
light is passed through the instrument, then $S = I = 1$, and the
cross talk term vanishes. We are then left with:
\begin{equation} \mathbf{M}_\mathrm{continuum} = \widehat{w}|_{d_0} / \widehat{w}|_{0}, \end{equation}
which describes the interferometer comb. As expected, the
properties of the comb are determined purely by the response
function, as discussed in section \ref{sec:comb}. There the comb was
described first for a delta-function response function, and then
for a top hat; the equation here represents the generalization to
any shape of response function.

Rewriting the first expression on the right hand side of equations
\ref{eqn:star1} and \ref{eqn:iodine1} in terms of $S$ and $I$ and
substituting into equation \ref{eqn:M1}, we can write:
\begin{equation}\label{eqn:exactsolution}
 \mathbf{M}\quad = \quad K_\mathrm{s}\mathbf{S} + K_\mathrm{i}\mathbf{I} \quad+\quad
 \frac{-\widehat{w}|_{d_0}+\widehat{siw}|_{d_0}} {\widehat{Sw}|_0+\widehat{Iw}|_0-\widehat{w}|_0+\widehat{siw}|_0},
\end{equation}
where the scalar quantities $K_\mathrm{s}$ and $K_\mathrm{i}$ are given by:
\begin{equation} \label{eqn:visscaling}
K_\mathrm{s} \equiv \frac{\widehat{Sw}|_0}{\widehat{Sw}|_0 +
  \widehat{Iw}|_0-\widehat{w}|_0+\widehat{siw}|_0}, \qquad
K_\mathrm{i} \equiv \frac{\widehat{Iw}|_0}{\widehat{Sw}|_0 +
  \widehat{Iw}|_0-\widehat{w}|_0+\widehat{siw}|_0}.
\end{equation}
Hence we see that we can now represent the combined star/reference data in terms
of a linear combination of the measured star and reference templates,
along with an error term.

The fraction on the right in equation \ref{eqn:exactsolution} contains
two terms in the numerator, the comb term, $\widehat{w}|_{d_0}$, and a
cross talk term, $\widehat{siw}|_{d_0}$. It is in principle possible
to arrange the instrument such that at delay $d=d_0$ the
interferometer comb has zero visibility, by choosing the delay and
slit width so that $\mathbf{M}_\mathrm{continuum}$ is at a zero point
of $\widehat{w}$ (see section \ref{sec:comb}). Alternatively, it is
possible to low-pass Fourier filter the data image before measuring
the whirls, essentially simulating a lower spectrograph resolution. In
either case, we assume that $\widehat{w}|_{d_0} \rightarrow 0$. If we
now also neglect all the cross talk terms $si$ following from equation
\ref{eqn:specexpanded}, we finally have the whirl addition
approximation, which we can write as
\begin{equation} \label{eqn:additionapprox}
  \boxed{\mathbf{M} \approx K_\mathrm{s}\mathbf{S} + K_\mathrm{i}\mathbf{I}.}
\end{equation}
$K_\mathrm{s}$ and $K_\mathrm{i}$ represent scaling factors in the
absolute visibilities of the two templates. In the case that we take
our normalization functions ($C_\mathrm{m}, C_\mathrm{s}$, and
$C_\mathrm{i}$) to be continuum normalization functions, then
remembering that the evaluation of a Fourier transform at $d=0$
represents the total integrated area under the function, we can try to
gain a handle on the expected sizes of these scaling factors. To the
extent that the total area under $Sw$ and $Iw$ is not much less than
that under $w$ (i.e. that the area in discrete absorption lines is
small, or $\int_{\Delta w} s\,\mathrm{d}\lambda \ll 1$ and
$\int_{\Delta w}i\,\mathrm{d}\lambda\ll 1$, where $\Delta w$ is a
representative width of the response function), equation
\ref{eqn:visscaling} implies that $K_\mathrm{s}, K_\mathrm{i} \approx
1$. This can easily be seen by rewriting in terms of $s$ and $i$
alone: we can then assume all the terms $\widehat{sw}|_0$,
$\widehat{iw}|_0$, $\widehat{siw}|_0 \ll 1$ --- the last because both
$s$ and $i$ are everywhere less than one by definition and so $si$
must always be even smaller than either --- and we find we are then
left with $K_\mathrm{s} \approx \widehat{w}|_0 / \widehat{w}|_0 = 1$,
and likewise for $K_\mathrm{i}$.

As far as the addition approximation holds good, and to the extent
that $K_\mathrm{s}$ and $K_\mathrm{i}$ are approximately constant
across all channels $j$, it is then a simple matter to allow for
Doppler and instrument drift by allowing the template whirls to rotate
in phase and translate in the dispersion direction as a function of
$j$; allowing $K_\mathrm{s}$ and $K_\mathrm{i}$ to vary as free
parameters as well, we can minimize $\chi^2$ in the residuals to find
the best fit solution compared to the measured data $\mathbf{M}$ for
the complete ensemble of wavelength channels. The difference between
the phase rotation of the star and that of the iodine (remembering to
account for wavelength dependence as necessary) yields the intrinsic
differential stellar Doppler shift, while the shifts in the dispersion
direction allow for Doppler shift of the stellar lines and any
instrumental image drift on the detector.

By these definitions, however, there is in fact little reason to assume that
$K_\mathrm{s}$ and $K_\mathrm{i}$ should be constant from channel to
channel. Furthermore, an iodine cell reference typically absorbs a
total of $\sim 40\%$ of the incident light, so that the assumption of
small area within the absorption lines is not necessarily robust
across the whole spectrum. Inspecting the terms in a little
more detail, we can recast them, rewriting
equation \ref{eqn:visscaling} as:
\begin{equation} 
K_\mathrm{s} = \frac{\widehat{Sw}|_0}{\widehat{Mw}|_0} =
\frac{\mathcal{F}[(\mathcal{S}/C_\mathrm{s})w]_0}{\mathcal{F}[(\mathcal{M}/C_\mathrm{m})w]_0}
=
\frac{C_\mathrm{m}}{C_\mathrm{s}}\frac{\widehat{\mathcal{S}w}|_0}{\widehat{\mathcal{M}w}|_0},
\end{equation}
and likewise for $K_i$ so that we have:
\begin{equation}\label{eqn:visscaling2}\boxed{
    K_\mathrm{s} =
    \frac{C_\mathrm{m}}{C_\mathrm{s}}\frac{\widehat{\mathcal{S}w}|_0}{\widehat{\mathcal{M}w}|_0},
    \qquad K_\mathrm{i} = \frac{C_\mathrm{m}}{C_\mathrm{i}}\frac{\widehat{\mathcal{I}w}|_0}{\widehat{\mathcal{M}w}|_0}.
  }\end{equation}
The terms are now written in terms of measurable quantities,
namely the total fluxes in each channel $j$ for the templates and the
data. We also see that they are dependent on the definition of the
functions $C_\mathrm{m}, C_\mathrm{s}$, and $C_\mathrm{i}$. Continuum
normalization functions could be determined by simply fitting a smooth
continuum function to the measured fluxes. There is, however, nothing in the
preceding analysis that {\em requires}
that $C_\mathrm{m}, C_\mathrm{s}$, and $C_\mathrm{i}$ be continuum
functions. Defining them as such allows for an
intuitive approach to visualize the effect of absorption lines, but they can in
fact be any function, subject only to our requirement that the
fractional deviations of the spectra from these functions (as
represented by $s$ and $i$) remain small, so that the cross-talk term
also remains small. It is arguably more appropriate to define the
functions to represent the {\em mean flux} across each of their
respective wavelength channels: in this case we see that
$K_{\mathrm{s}}$ and $K_{\mathrm{i}}$ simplify immediately to exactly
unity, independent of wavelength channel, so that they drop out of
equations \ref{eqn:exactsolution} and \ref{eqn:additionapprox}. The
difference is absorbed in the cross talk-term through its dependence
on $s$ and $i$, which in turn are also dependent on $C_{\mathrm{s}}$
and $C_{\mathrm{i}}$ respectively. Written in this way, the whole of
the addition approximation error is included in the single cross-talk term,
$\widehat{siw}|_{d_0}$, in equation \ref{eqn:exactsolution}.

We now have an approximate formalism for solving for stellar Doppler
shifts from combined star/reference data, where the reference spectrum
multiplies the stellar spectrum. The above analysis is only useful,
however, in as far as the approximation that the cross talk, $si$, is
very small holds well. It appears, however, that as it stands, this
approximation is in fact not accurate enough for exoplanet
searches. In section \ref{sec:additionapproxerr}, we derive an
estimate of the errors resulting from the approximation, and find that
systematics as large as $50\ms$ or more can arise. Clearly this cannot
be neglected. Approaches to correcting or avoiding the error are
discussed in section \ref{sec:discussion}.


\subsubsection{An Alternative: Combined-beam Reference} \label{sec:addedref}

One possible solution to the problem of the addition approximation is
to actually physically superpose a reference spectrum on top of the
stellar target spectrum, for example by splicing two input fibers into
one, one coming from the telescope and one from the reference lamp. In
this case, the two spectra now combine additively instead of
multiplicatively. We can then write:
\begin{equation}
\mathcal{M}  =  A_\mathrm{s} \mathcal{S} + A_\mathrm{i} \mathcal{I} 
\end{equation}
where $A_\mathrm{s}$ and $A_\mathrm{i}$ are scaling factors to allow for flux
differences between the templates and data (note that two such factors
are now required).
Once again, following equation \ref{eqn:cplxvis} we can now write:
\begin{eqnarray}\label{eqn:additiondata1}
\mathbf{M} &=& \frac{\mathcal{F}[\mathcal{M}w]_{d_0}}{\mathcal{F}[\mathcal{M}w]_{0}}\nonumber\\
&=&\frac{\mathcal{F}[(A_\mathrm{s}\mathcal{S}+A_\mathrm{i}\mathcal{I})w]_{d_0}}{\mathcal{F}[(A_\mathrm{s}\mathcal{S}+A_\mathrm{i}\mathcal{I})w]_{0}}\nonumber\\
&=&\frac{A_\mathrm{s}\widehat{\mathcal{S}w}|_{d_0}+A_\mathrm{i}\widehat{\mathcal{I}w}|_{d_0}} 
        {A_\mathrm{s}\widehat{\mathcal{S}w}|_0+A_\mathrm{i}\widehat{\mathcal{I}w}|_0},
\end{eqnarray}
or alternatively,
\begin{equation}\label{eqn:additiondata2}
\boxed{\mathbf{M}=K^\prime_\mathrm{s}\mathbf{S} + K^\prime_\mathrm{i}\mathbf{I},}
\end{equation}
where we define:
\begin{equation}
  K^\prime_\mathrm{s} \equiv \frac{A_\mathrm{s}\widehat{\mathcal{S}w}|_0}
  {A_\mathrm{s}\widehat{\mathcal{S}w}|_0+A_\mathrm{i}\widehat{\mathcal{I}w}|_0},\qquad
  K^\prime_\mathrm{i} \equiv \frac{A_\mathrm{i}\widehat{\mathcal{I}w}|_0}
  {A_\mathrm{s}\widehat{\mathcal{S}w}|_0+A_\mathrm{i}\widehat{\mathcal{I}w}|_0},
\end{equation}
or:
\begin{equation}\boxed{
  K^\prime_\mathrm{s} = A_\mathrm{s}\frac{\widehat{\mathcal{S}w}|_0}
  {\widehat{\mathcal{M}w}|_0},\qquad
  K^\prime_\mathrm{i} = A_\mathrm{i}\frac{\widehat{\mathcal{I}w}|_0}
  {\widehat{\mathcal{M}w}|_0}.
}\end{equation}
We see that we now have an exact expression for $\mathbf{M}$, with the difference being
that we now need to take into account the flux scaling factors $A_\mathrm{s}$
and $A_\mathrm{i}$, where previously the flux scaling factor had canceled.

There is also a constraint on the visibility scaling constants
$K^\prime_\mathrm{i}$ and $K^\prime_\mathrm{s}$. Since the Fourier
transforms at $d=0$ represent total fluxes within the channel, flux
conservation means that we can write:
\begin{equation}
   A_\mathrm{s}\widehat{\mathcal{S}w}|_0 +
   A_\mathrm{i}\widehat{\mathcal{I}w}|_0
   = \widehat{\mathcal{M}w}|_0.
\end{equation}
Dividing through by the flux in the combined star/iodine data,
$\widehat{\mathcal{M}w}|_0$, and substituting the visibility scaling
constants, we find:
\begin{equation}
K^\prime_\mathrm{i} + K^\prime_\mathrm{s} = 1
\end{equation}

As before, we can solve for phase rotation and dispersion shift by
$\chi^2$ minimization, this time additionally solving for the two flux
scaling constants. It is interesting to note that if we multiply
through both sides of equation \ref{eqn:additiondata2} by the
denominator, $ {\widehat{\mathcal{M}w}|_0}$ (which represents the
total flux along the channel in the combined data), we essentially
find we have an expression which is a summation of flux $\times$
visibility terms. Since visibility is defined as
$(I_\mathrm{max}-I_\mathrm{min})/(I_\mathrm{max}+I_\mathrm{min})$,
where $I_\mathrm{max}$ and $I_\mathrm{min}$ are the maximum and
minimum fringe intensities, then multiplying by total flux in the
channel gives a quantity equal to the amplitude of the fringe. Hence
equation \ref{eqn:additiondata2} is really simply summing fringe
amplitudes, and is exactly what we expect when the two input spectra
are combined additively: the resulting image on the detector should
simply be a direct flux summation of the respective images that would
be obtained individually.

%% file: section3.tex
\section{Sources of Error} \label{sec:CHerrors}

Here we provide derivations of some useful formulae for estimating the
errors from certain sources for which we have been able to find
analytical approaches. These include photon errors; additive spectral
contamination errors, such as moonlight background, crowded targets,
etc.; and multiplicative fringe-visibility contamination errors, which
include in particular the cross-talk error due to the whirl addition
approximation for in-beam absorption reference sources, but which can
also be applied to other effects such as residual interferometer comb
(again, in the case of an in-beam reference). The latter formulae are
potentially applicable to a number of different error sources, and all
are likely to be useful for any implementation of a DFDI
instrument.

Since this is primarily a theory paper, we do not attempt to provide a
comprehensive accounting of error sources: many are
instrument implementation specific, or data reduction pipeline
specific, and better suited to empirical or semi-empirical assessment
through simulations and experimentation. Such work is still ongoing
with the ET project. For more complete discussion of specific errors
in the ET project, we point the reader to upcoming MARVELS
publications on the instrument (J. Ge et al. 2010, in preparation) and pipeline
(B. Lee et al. 2010, in preparation); more detailed discussions of errors from
earlier ET work can also be found in \citet{MyThesis} and
\citet{suvrathsthesis}. Table \ref{tbl:errors} provides a summary of
the examples of applications of the error formulae provided in the text.

\begin{deluxetable}{lcccl}
\tablecaption{Summary of Example Error Magnitudes\label{tbl:errors}}
\tablewidth{0pt}
\tablehead{ \colhead{Noise Source} &
\colhead{Subsection} & \colhead{Approx. Magnitude} &
 \\ & & \colhead{$(\ms)$}
 }
\startdata

Photon shot noise -- multiplied ref.\tablenotemark{a}& \ref{sec:normalphotonerr} & 3.2
\\
Photon shot noise -- added ref.\tablenotemark{a}& \ref{sec:addedspecpherr} & 3.6
\\
Photon shot noise -- separate ref.\tablenotemark{a} & \ref{sec:separaterefpherr} & 2.9
\\
Moonlight contamination & \ref{sec:mooncontamination} & $\lesssim 41$
\\
Residual interferometer comb\tablenotemark{b} & \ref{sec:comberr} & 9
\\
Addition approximation\tablenotemark{b} & \ref{sec:additionapproxerr}
& 50
\\

\enddata
\tablecomments{Error magnitudes as calculated in the text are
  listed: these are examples for illustration only, and each is
  highly variable and dependent on specific circumstances. See the text for
  assumptions made in each case.}
\tablenotetext{a}{Assuming iodine reference -- see text for
  improvements using ThAr in the added-reference case.}
\tablenotetext{b}{Applies only for
  multiplied reference spectrum.}
\end{deluxetable}


\subsection{Photon Errors}\label{sec:photonerrors}

The errors due to photon shot noise provide an important baseline for
any instrument.  They indicate the absolute limit to the precision
that can be achieved, and drive throughput and (for DFDI instruments)
fringe visibility considerations for the optical design. It is
entirely reasonable to conceive of a photon-limited DFDI-type
instrument. However, even in cases where photon noise is dominated by
other effects in the very high precision regime, photon noise
inevitably becomes significant at the faint end of the stellar target
sample. For the MARVELS/Keck ET, geared toward moderate precision surveys of
fainter targets, although other errors dominate the instrument
requirements error budget at the brightest ($V\sim 8\vmag$) end of the
target range, photon noise becomes a significant part of the
error at fainter levels (down to $V=12\vmag$, $\sim 21.5\ms$ of a
total $35.0\ms$ -- see \citet{GeMARVELSWhitePaper}). In the high precision,
high-flux regime, (e.g., a planned $1\ms$--level cross-dispersed DFDI
upgrade for the KPNO ET), the photon error is also important as it
indicates the level below which other sources of systematic and random
error must be driven.

The photon error in the phase measurement (and hence velocity
measurement) from a single channel can be estimated following
\cite{Ge2002}. This gives essentially
\begin{equation}\label{eqn:Ge2002pherr}
  \varepsilon_{v,j} \approx \quad \frac{1}{\pi\surd{2}} \, \frac{c\lambda}{d
    \alpha_j \surd F_{j}} \quad  = \quad
  \Gamma\frac{\surd{2}}{\alpha_j\surd{F_j}} 
\end{equation}
where $\varepsilon_{v,j}$ is the error in velocity due to channel $j$
alone, $c$ is the speed of light, $\lambda$ is the wavelength, $d$ is
the optical delay, $\alpha_j$ is the visibility of the fringe, $F_{j}$
is the total flux in the channel, and $\Gamma$ is the usual
phase-velocity scaling factor (equation \ref{eqn:Gamma}, ignoring the
negative sign since we are interested only in the
magnitude).\footnote{The small difference in the numerical factor in
  the denominator of equation \ref{eqn:Ge2002pherr} ($\pi\surd{2}$
  versus 4) is due to using the rms slope of the fringe, rather than the
  mean absolute slope used in \cite{Ge2002}. Monte Carlo simulations
  of sinusoid fits suggest that the rms slope gives more accurate
  results.} The terms following the $\Gamma$ represent the error in
phase due to the photon noise, $\varepsilon_{\phi,j} =
\surd{2}/(\alpha_j\surd F_j)$. Following a similar derivation, it is
straightforward to show that the error in {\em visibility} due to
photon noise, $\varepsilon_{\alpha,j}$, is given by
\begin{equation}
\varepsilon_{\alpha,j} = \sqrt{\frac{2}{F_j}},
\end{equation}
and hence, assuming independent errors, there is a useful simple
relationship between the errors in phase and visibility:
\begin{equation}\label{eqn:phaseerrvsviserr}
\varepsilon_{\phi,j} = \frac{\varepsilon_{\alpha,j}}{\alpha_j}.
\end{equation}

As a general rule, we can see from equation \ref{eqn:Ge2002pherr} that precision goes
with the inverse root of flux, as one would expect, and also as the
inverse of visibility: higher flux and/or higher visibility mean
better precision. From this formula we can derive the photon errors in
the final differential RV for different calibration scenarios.

For simplicity in the following formulae, we take $\lambda$ to be
constant, taking the wavelength value at the center of the spectrum,
since it varies by only $\sim 10\%$ from one end of the spectrum to
the other in the current ET instruments. For an instrument with a very
large bandwidth, however, it may be necessary to consider it properly
as a function of channel, $\lambda_j$. This simply means it cannot be
taken outside the brackets as in the following derivations, but
otherwise the formalism is the same.

\subsubsection{Photon Error for Multiplied Reference}\label{sec:normalphotonerr}

To calculate the expected error in an RV measurement for a
single data frame, assuming an instrument configuration where an
iodine or other reference spectrum multiplies the input stellar
spectrum, we consider the resulting data spectrum as consisting of two
components, a star component, and an iodine component. The calculated
phase shift due to intrinsic target Doppler shift, $\Delta \phi$ is
given by:
\begin{equation} \label{eqn:phaseshiftcomponents}
  \Delta \phi=\langle \phi_{\mathrm{sm},j} - \phi_{\mathrm{st},j}\rangle \: - \: \langle\phi_{\mathrm{im},j} - \phi_{\mathrm{it},j}\rangle,
\end{equation}
where $\langle\dots\rangle$ here represents a {\em weighted} mean over
all $j$, $\phi_{\mathrm{sm},j}$ and $\phi_{\mathrm{im},j}$ represent
the phases for the star and iodine components of the combined
star/iodine data (`measurement') frame, and $\phi_{\mathrm{st},j}$ and
$\phi_{\mathrm{it},j}$ are the phases measured in the separate pure
star and iodine templates. For convenience, we immediately map these
phases to corresponding `velocity' measurements by multiplying both
sides by $\Gamma$ to give a velocity shift, $\Delta v$ (though with
the caveat that a velocity measurement of a single channel in a single
spectrum has no physical meaning in itself until it is differenced
with another spectrum):
\begin{equation} \label{eqn:velshiftcomponents}
  \Delta v=\langle v_{\mathrm{sm},j} - v_{\mathrm{st},j}\rangle - \langle v_{\mathrm{im},j} - v_{\mathrm{it},j}\rangle,
\end{equation}
Using $\varepsilon_{v}$ with corresponding subscripts to represent the
various errors in this equation, we can expect a total
photon error in $\Delta v$ to be given by:
\begin{equation} \label{eqn:totalpherr}
  \varepsilon_{v}^2 = \left[E_j\left(\sqrt{\varepsilon_{v,\mathrm{sm},j}^2 + \varepsilon_{v,\mathrm{st},j}^2}\,\right)\right]^2 + \left[E_j\left(\sqrt{\varepsilon_{v,\mathrm{im},j}^2 + \varepsilon_{v,\mathrm{it},j}^2}\,\right)\right]^2,
\end{equation}
where $E_j(\sigma)$ represents the standard statistical
error in a weighted mean:
\begin{equation}\label{eqn:errcombine}
  E_j(\sigma_j) \equiv \frac{1}{\sqrt{\sum_{j} 1/\sigma_j^2}}.
\end{equation}

In practice, the two template terms in equation \ref{eqn:totalpherr}
are neglected, for two reasons. The first is simply because in general
the templates will have significantly higher flux than the data frame:
the iodine template can be taken with arbitrarily high flux since it
is obtained with a quartz lamp as a source; and the stellar template
is usually deliberately taken with higher flux than the data so that
it does not compromise the entire data set. The second reason is a
little more subtle. All RV measurements with this kind of instrument
are differential, measured relative to the two templates which
effectively set the zero point of the measurements for the star and
iodine, as seen in equation \ref{eqn:velshiftcomponents}. Since this
`zero point' is the same for every RV measurement, any error in the
zero point will not contribute to the rms scatter in a set of
measurements which uses the same templates.

This last statement holds true to a point: accuracy in the
templates is still needed in order to disentangle the stellar and
iodine components of the combined data. From simulations of ET
fringing spectra, we find, for example, that for a multiplied iodine
reference, using a G0 or G2V stellar template in place of a G8V
template yields an rms error of $11\ms$ over large ($60\kms$)
differential velocity shifts. (Depending on the precision required,
this points toward the interesting possibility of using templates of
different stars from the target star: this could allow, for example,
for higher S/N templates when observing very faint targets, or perhaps
for disentangling the signals from double-lined spectroscopic
binaries.)

Since photon errors go as
$1/\surd{\mathrm{flux}}$, the remaining terms,
$E_j(\varepsilon_{v,\mathrm{sm}})$ and $E_j(\varepsilon_{v,\mathrm{im}})$, can be
estimated by scaling the respective template errors (which, unlike the
measurement component errors, can be determined directly
from equation \ref{eqn:Ge2002pherr}) by the flux difference between
the templates and data, giving:
\begin{eqnarray} 
  \varepsilon_{v}^2 & = & [E_j(\varepsilon_{v,\mathrm{sm},j})]^2 +
  [E_j(\varepsilon_{v,\mathrm{im},j})]^2 \label{eqn:basicpherr}\\
  & \approx & \frac{\overline{F}_{\mathrm{st}}}{\overline{F}_{\mathrm{m}}}\,[E_j(\varepsilon_{v,\mathrm{st},j})]^2
        +
        \frac{\overline{F}_{\mathrm{it}}}{\overline{F}_{\mathrm{m}}}\,[E_j(\varepsilon_{v,\mathrm{it},j})]^2 \label{eqn:intermediatepherr}
\end{eqnarray}
where $\overline{F}_{\mathrm{st}}$, $\overline{F}_{\mathrm{it}}$, and
$\overline{F}_{\mathrm{m}}$ represent the mean fluxes across the whole
star template, iodine template and data frame respectively.
Explicitly substituting equation \ref{eqn:Ge2002pherr} into equation
\ref{eqn:intermediatepherr}, we find:
\begin{equation} \label{eqn:finalphotonerror}
\boxed{
  \varepsilon_{v} = 
  \Gamma\surd{2}\;
  \sqrt{
    \frac{\overline{F}_{\mathrm{st}}}{\overline{F}_{\mathrm{m}}}
    \left [E_j \left
        (\frac{1}{\alpha_{\mathrm{st},j}\surd{F_{\mathrm{st},j}}}
      \right )
    \right ]^2
    +
    \frac{\overline{F}_{\mathrm{it}}}{\overline{F}_{\mathrm{m}}}
    \left  [E_j \left
        (\frac{1}{\alpha_{\mathrm{it,j}}\surd{F_{\mathrm{it},j}}}
      \right )
    \right ]^2
  },
}
\end{equation}
where the error combination function, $E_j$, is given by equation \ref{eqn:errcombine}.

Hence we have a quadrature summation of the photon errors due to the
star and reference components of the combined star/iodine data, each
being the weighted expected error in velocity across the respective
template spectra scaled to the flux level of the data. As one would
expect, the error goes with the inverse root of the mean flux in the
data spectrum, $\left(\overline{F_{\mathrm{m}}}\right)^{-1/2}$; the
error in each of the two components will also scale as the inverse of
the visibility in the respective fringing spectra. Written in this
form, the $E(\dots)$ terms need only be calculated once, representing
photon errors for each template: they then can be conveniently scaled
and combined to give the error in each data frame for the source.

We note that these formulae for the photon limit are for the values
expected {\em given} the fringe visibility that was obtained. Various
instrument effects -- for example defocus, or a non-optimal delay for
the stellar line width -- can reduce the visibility from its optimum
and hence reduce this photon limiting precision.

\begin{figure*}[tbp]
  \epsscale{0.7}
  \plotone{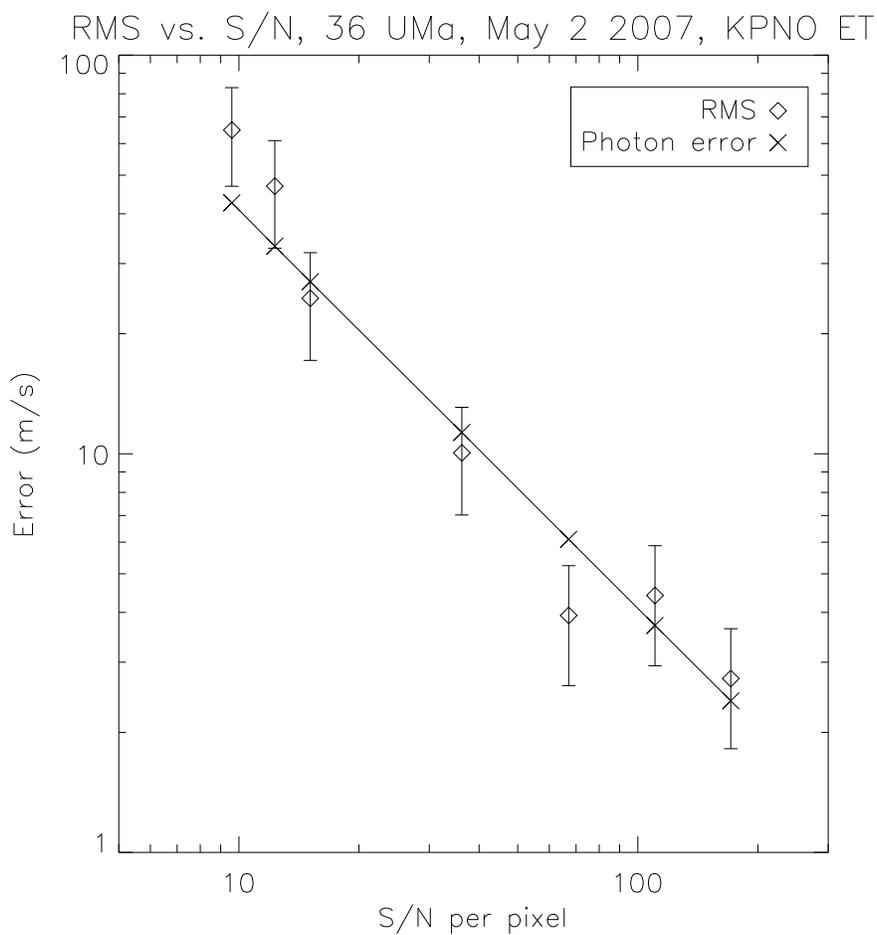}
  \caption{\label{fig:rmsvssn}Measured rms vs. S/N per pixel for
    the bright stable star 36~UMa over a total of approximately
    2\,hr. Obtained with the KPNO ET on May 2, 2007, with varying
    exposure length to achieve different S/N levels, (approx. 5--6
    data points per S/N level). Diamonds indicate the rms, with error
    bars corresponding to the uncertainty due to the number of data
    points over which the rms is calculated. Crosses and line indicate
    the corresponding calculated photon limit.}
\end{figure*}

This has been the formalism employed in calculating the photon error
for the Kitt Peak single-object ET for operations with an in-beam
iodine cell. As an example, an observation taken at very high flux
with the KPNO 2.1m ET run in May 2007 of 36~UMa (stable,
$V=4.84\vmag$, $10\min$ exposure) gives mean signal/noise ratios (S/N)
per pixel for star template, iodine template, and data frame of 222,
146 and 179 respectively. These values give photon errors for the star
and iodine components of 2.8 and $1.5\ms$ respectively, which when
added in quadrature give a total photon error of $3.2\ms$. The KPNO
instrument design is such that both output beams from the Michelson
interferometer are recovered, and this result is for only one of the
two beams. Averaging over the two beams therefore in fact gives a
further improvement of $1/\surd 2$ in photon precision; for
simplicity, and for comparison with the following sections, we
consider only one beam here, however. It is interesting to note that
the error due to the iodine reference is in fact comparable in
magnitude to that due to the star, since the signal in the iodine
component of the data frame is intrinsically limited by the magnitude
of the target being observed. Figure \ref{fig:rmsvssn} shows a
comparison of the actual rms (on the very short term) with the
calculated photon errors using this formalism, obtained with the KPNO
ET on the bright stable star 36~UMa over a total of $\sim2$\,hrs,
showing good agreement. The preceding example calculation is based on
a data point at the high-flux end of this data set. (For the purposes
of the plot, the two interferometer output beams are averaged.)

These calculations assume that the flux ratio terms remain the same
from channel to channel, so that an overall mean scaling can be
applied. This is not strictly accurate (e.g., if line depths are very
deep and broad, or the pure star and pure iodine continuum functions
are very different), but is taken to work to a reasonable
approximation, and seems to correspond quite well with real results.
In the event that a more accurate calculation is needed, however, it
is a simple enough matter to introduce channel-dependent flux ratios
for each element $j$ within the summations.


\subsubsection{Photon Error for Added Reference Spectrum} \label{sec:addedspecpherr}

In the case of the reference spectrum being combined additively,
rather than multiplicatively, the photon errors must be calculated
differently. However, we can follow a somewhat similar
approach. Again, we consider the errors due to star and iodine
components of the combined star/iodine data, and neglect the errors
due to the templates, so that, as for equation \ref{eqn:basicpherr}:
\begin{equation} \label{eqn:totalpherr_addition}
  \varepsilon_{v}^2 = [E_j(\varepsilon_{v,\mathrm{sm},j})]^2 + [E_j(\varepsilon_{v,\mathrm{im},j})]^2,
\end{equation}
where $E_j$ is again defined as in equation \ref{eqn:errcombine}. The
individual components $\varepsilon_{v,\mathrm{sm},j}$ and
$\varepsilon_{v,\mathrm{im},j}$ must be reevaluated, however, since the
photon noise from the two separate sources will now combine additively
(for example, if one of the sources is considerably brighter than the
second, its photon noise will dominate over the signal in the
second). We can think of an effective visibility for the two
components in the combined data, $\alpha_{\mathrm{sm},j}$ and
$\alpha_{\mathrm{im},j}$. Remembering that fringe amplitude is given
by the product of the visibility and the mean flux in the fringe, we
can write
\begin{equation}
\alpha_{\mathrm{sm},j}\overline{F_{\mathrm{m},j}} =
\alpha_{\mathrm{st},j} A_{\mathrm{s}} \overline{F_{\mathrm{st},j}}
\quad ; \quad
\alpha_{\mathrm{im},j}\overline{F_{\mathrm{m},j}} =
\alpha_{\mathrm{it},j} A_{\mathrm{i}} \overline{F_{\mathrm{it},j}},
\end{equation}
where $\alpha_{\mathrm{st},j}$ and $\alpha_{\mathrm{it},j}$ are the
fringe visibilities for channel $j$ in the star and iodine templates
respectively; and ${\overline{F_{\mathrm{st},j}}}$,
${\overline{F_{\mathrm{it},j}}}$, and ${\overline{F_{\mathrm{m},j}}}$
are the mean fluxes across the channel for the star template, iodine
template, and data (measurement) frame respectively. $A_{\mathrm{s}}$
and $A_{\mathrm{i}}$ are wavelength-independent scaling factors that
allow for flux differences between the templates and the respective
data components, as in section \ref{sec:addedref}. Hence we find
\begin{equation}
  \alpha_{\mathrm{sm},j} = \frac{\alpha_{\mathrm{st},j} A_{\mathrm{s}} F_{\mathrm{st},j}} {F_{\mathrm{m},j}} \quad ;\quad
  \alpha_{\mathrm{im},j} = \frac{\alpha_{\mathrm{it},j} A_{\mathrm{i}} F_{\mathrm{it},j}} {F_{\mathrm{m},j}},
\end{equation}
where $F_{\mathrm{st},j}$ and ${F_{\mathrm{it},j}}$ are the
total fluxes across the channel for the star and iodine templates, and $F_{\mathrm{m},j}$ is
the total flux in the channel for the data
frame. Substituting these effective visibilities in equation
\ref{eqn:Ge2002pherr} gives:
\begin{eqnarray}
  \varepsilon_{v,\mathrm{sm},j} & = & \Gamma\surd 2 \, \frac{\surd{F_{\mathrm{m},j}}}{\alpha_{\mathrm{st},j}A_\mathrm{s} F_{\mathrm{st},j}} \nonumber \\
  \varepsilon_{v,\mathrm{im},j} & = & \Gamma\surd 2 \, \frac{\surd{F_{\mathrm{m},j}}}{\alpha_{\mathrm{it},j}A_\mathrm{i} F_{\mathrm{it},j}}.
\end{eqnarray}
Using these we can now evaluate equation \ref{eqn:totalpherr_addition}
to obtain an estimate of the photon limiting error, so that:
\begin{equation}\label{eqn:addedpherr}
  \boxed{
    \varepsilon_{v} =
    \Gamma\surd{2}\;
    \sqrt{
      \left[E_j\left(\frac{\surd{F_{\mathrm{m},j}}}{\alpha_{\mathrm{st},j}A_\mathrm{s}F_{\mathrm{st},j}}\right)\right]^2
      +
      \left[E_j\left(\frac{\surd{F_{\mathrm{m},j}}}{\alpha_{\mathrm{it},j}A_\mathrm{i}F_{\mathrm{it},j}}\right)\right]^2 },
  }
\end{equation}
where, due to flux conservation,  $A_\mathrm{s}$ and $A_\mathrm{i}$ are subject to the
constraint:
\begin{equation}\label{eqn:addedpherr2}
A_{\mathrm{s}} F_{\mathrm{st},j} + A_{\mathrm{i}} F_{\mathrm{it},j} = F_{\mathrm{m},j}.
\end{equation}

Again we have found a quadrature summation of errors due to the star
and reference components, scaled to match the respective component
fluxes in the data, very similar to equation
\ref{eqn:finalphotonerror}. However, in this case, the scaling factors,
$A_{\mathrm{s}}$ and $A_{\mathrm{i}}$ must be determined as parameters
during the velocity shift solution, and $A_{\mathrm{s}} F_{\mathrm{st},j}$
and $A_{\mathrm{i}} F_{\mathrm{it},j}$ represent the fluxes in the
star and iodine components of the data, respectively. 

This time, we do not attempt to assume channel-independent flux
ratios. This is because for additively combined references it becomes
possible to consider using emission spectra (e.g., a ThAr lamp) as the
reference, rather than the usual iodine absorption spectrum. Clearly
the flux ratio between data and reference template frames is very
different for regions where there are no reference emission lines
compared to those where emission lines are present. It is therefore
not reasonable to take the flux terms outside the summation in the
error combination function $E_j$.

To gain a handle on the behavior of equation \ref{eqn:addedpherr}, we
can see that if we consider only a single channel, so that for a
function $f$, $E(f)\rightarrow f$, and assume both that the source
and reference visibilities are roughly equal (reasonable for star and iodine,
to order of magnitude) and that the total flux $F_{\mathrm{m}}$
remains constant, then to minimize the total error, we need only to
minimize the function:
$(A_{\mathrm{s}}F_{\mathrm{st},j})^{-2} + (A_{\mathrm{i}}F_{\mathrm{it},j})^{-2}.$
Given the constraint of equation \ref{eqn:addedpherr2} it is
straightforward to show that this is minimized when
$A_{\mathrm{s}}F_{\mathrm{st},j} = A_{\mathrm{i}}F_{\mathrm{it},j}$,
in other words, when the component star and iodine fluxes are
approximately equal. When either component has a very small flux
compared to the other, one or other of the terms in equation
\ref{eqn:addedpherr} will become very large. Broadly speaking, then,
we can see that the fluxes of star and reference need to be balanced
in order to minimize photon error.

In practice, rather than the total flux being constant, it is of more
interest to hold the stellar component constant and vary the reference
component to find the optimum; using real spectra, and allowing
differing visibilities, the balance point becomes a little skewed from
unity. The exact optimal balance point depends on the spectra in
question. Allowing for the gain in optical throughput from losing the
absorption in the gas cell reference, this equation at its balance
point generally gives photon errors on a similar level to the photon
errors for a multiplied iodine reference, if we use iodine spectra as
references in both cases (i.e., tungsten-illuminated iodine in the
added-reference scheme). Using the same observations as in
section \ref{sec:normalphotonerr} to calculate error estimates as if
the spectra had been added, and assuming that the flux level of the
star in the template and the hypothetical combined observation is the
same, we find an optimal ratio of iodine to star flux of 0.96 and a
total photon error of $3.6\ms$, comprised of iodine and star component
errors of $3.1\ms$ and $1.8\ms$, compared to the total error of
$3.2\ms$ for multiplied spectra. That the two are similar is not
surprising: adding a reference spectrum to the stellar spectrum at a
matching flux level will approximately halve fringe visibility and
hence double the error, but also double the flux, reducing the error
by $1/\surd 2$, giving a total $\surd 2$ increase in the error
size. This coincidentally matches the increase in error size for
in-beam-iodine calibration due to the fact that the iodine typically
absorbs $\sim 50\%$ of the incident light. (The slight mismatch in the
figures calculated is due to the fact that in the multiplicative case,
the combined data frame actually had particularly high flux, probably
because of better sky transparency at the time the frame was taken
than when the template was taken). 

The above argument holds true for iodine since the continuum shape and
fringe visibilities are broadly similar to those of the stellar
spectrum. If we instead use a ThAr emission spectrum for the added
reference, we appear to perform even rather better than the in-beam
iodine case: the same calculations as above with a ThAr spectrum
replacing the iodine spectrum yield a total photon error of $2.5\ms$,
with star and ThAr components of $2.4\ms$ and $0.65\ms$
respectively (with an optimum ratio of mean fluxes of 0.26 -- now
substantially different because of the very different nature of an
emission spectrum). ThAr also shows a weaker dependence on relative flux level,
which gives it an advantage in terms of practical application since
less effort would need to be expended on matching the brightness to
each target observation. This is likely because most of
the Doppler information is primarily concentrated in a few bright
lines in the ThAr, where it is spread more broadly across the stellar
spectrum. Where the ThAr lines are strong, the stellar Doppler
information is likely largely lost due to the added photon
noise. However, since there are relatively few such lines, there is
not too much impact on the total stellar Doppler information, and
increasing the ThAr flux does not make as large a difference as for
the case of an iodine spectrum. We note, however, that at very high
flux levels, saturation of the brightest ThAr emission lines is likely
to complicate this analysis somewhat.

Added-beam reference calibration provides one possible solution to the
reference addition approximation error discussed in section
\ref{sec:additionapproxerr}, and the discussion here should provide a
formalism for calculating the photon errors. Such a calibration
approach, however, has not yet been attempted within the ET program,
although basic simulations bear out these calculations.

\subsubsection{Photon Error with a Separate Reference}\label{sec:separaterefpherr}

Finally we consider the simple case where there is no simultaneous
common-path reference, but rather a reference separated either
spatially or in time. Once again, we find a find a weighted mean velocity
shift between (now pure) star measurement and some reference star
template, and the same between a pure reference spectrum measurement
and a corresponding template. (The reference need not be iodine, but
we retain the `i' subscript notation for consistency). The results are
differenced to obtain a corrected intrinsic stellar Doppler shift.
If we neglect the template errors as before, then the photon
errors for the data frame are found again similarly to equation
\ref{eqn:basicpherr}:
\begin{equation}
  \varepsilon_{v}^2 = [E_j(\varepsilon_{v,\mathrm{s},j})]^2 +  [E_j(\varepsilon_{v,\mathrm{i},j})]^2.
\end{equation}
The difference is that here we use subscripts ``s'' and ``i,'' rather than
``sm'' and ``im,'' to indicate that we are no longer looking at components
of a combined reference/star measurement frame, but at pure star and
pure reference measurements respectively. Again substituting the basic
photon error equation, \ref{eqn:Ge2002pherr}, we obtain:
\begin{equation}\boxed{
  \varepsilon_{v} = 
  \Gamma\surd{2}\;
  \sqrt{
    \left [E_j \left
        (\frac{1}{\alpha_{\mathrm{s},j}\surd{F_{\mathrm{s},j}}}
      \right )
    \right ]^2
    +
    \left  [E_j \left
        (\frac{1}{\alpha_{\mathrm{i,j}}\surd{F_{\mathrm{i},j}}}
      \right )
    \right ]^2
  }.}
\end{equation} 

The form of the equation is now much simpler, since no flux or
visibility scaling is required. Errors again go as the inverse of the
visibilities of star and iodine, and as the inverse root of the flux.
It may also be the case (indeed, observations should be taken such that
it is the case) that the reference spectrum has significantly higher
S/N, and therefore its photon errors can be neglected, so that only
the first error combination term in the square root remains.

Taking our same data and templates once again, we can calculate a
hypothetical photon error for comparison: this time, using iodine as a
separate reference yields a total error of $2.9\ms$ comprising star
and iodine components of $2.3\ms$ and $1.9\ms$ respectively (note that
the iodine error level here is relatively high in comparison to the
star component: this is purely because of the exceptionally high flux
from the star in these particular observations); using ThAr instead
yields a total error of $2.4\ms$, with star and ThAr components of
$2.3\ms$ and $0.77\ms$ (though again we have not included the effects
of saturation in the ThAr calculation, which may increase the ThAr
errors somewhat).

This approach is appropriate to the MARVELS/Keck ET, where pure star
science exposures are bracketed in time with pure iodine reference exposures,
and the instrument is highly stabilized in both pressure and
temperature. The baseline design requirements anticipate a photon
error of $3.5\ms$ at $V=8\vmag$, and $21.5\ms$ at
$V=12\vmag$ \citep[see section \ref{sec:inpractice} and][]{GeMARVELSWhitePaper}.

\subsection{Additive Contaminating Spectra}\label{sec:contaminatingspectra}
It is often useful to be able to calculate a rough estimate of the
errors due to contaminating additive background spectra. We derive a
formalism for doing so here. This formalism will enable us to
calculate the effect of background moonlight contamination,
contaminating background stars, or double-lined spectroscopic
binaries, for example. In addition, we will then be able to extend the
formalism to treat multiplicative (i.e., flux independent)
contaminants, such as any residual unfiltered comb presence or the
iodine/star cross-talk term that causes the reference addition
approximation error, and try to assess their relative significance.

\subsubsection{Derivation}
\begin{figure}[tbp]
  \epsscale{0.5}
  \plotone{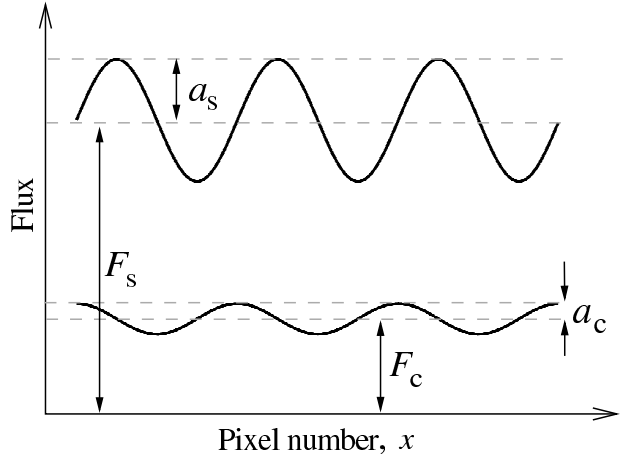}
  \caption[Fringe along one channel due to target (upper curve), and contaminating low
  flux fringe (lower curve)]{\label{fig:amplitudes}Fringe
  along one channel due to source (upper curve) and contaminating low
  flux fringe (lower curve). Measured fringe is a summation of these
  two fringes.}
\end{figure}

Figure \ref{fig:amplitudes} shows a fringe along one detector column
(in the slit direction) due to the target source alone, with fringe
amplitude $a_\mathrm{s}$, mean flux $F_{\mathrm{s}}$, and phase
$\phi_\mathrm{s}$. For simplicity we assume no iodine fiducial
reference, since we are only aiming for an order-of-magnitude
estimate. A second contaminating fringe of lower amplitude
$a_\mathrm{c}$ and mean flux $F_\mathrm{c}$ due to background
contamination is also shown, with phase $\phi_\mathrm{c}$. If the
spatial frequency of the fringes is $f$, then the summation of these
two fringes will give the total (also sinusoidal) measured
fringe:
 \begin{eqnarray} & F_{\mathrm{s}} + \Re\{a_\mathrm{s}
  e^{i(fx+\phi_\mathrm{s})}\} + F_\mathrm{c} +
  \Re\{a_\mathrm{c}
  e^{i(fx+\phi_\mathrm{c})}\} \nonumber \\
  = & F_{\mathrm{s}}+F_\mathrm{c}+\Re\{a_\mathrm{s}
  e^{i(fx+\phi_\mathrm{s})}+a_\mathrm{c}
  e^{i(fx+\phi_\mathrm{c})}\},
\end{eqnarray}
where $x$ identifies position along the slit. $F_{\mathrm{s}}+F_\mathrm{c}$ represents the
mean value of the measured flux. The last term represents the varying
sinusoidal net fringe.

\begin{figure}[tbp]
  \epsscale{0.8}
  \plotone{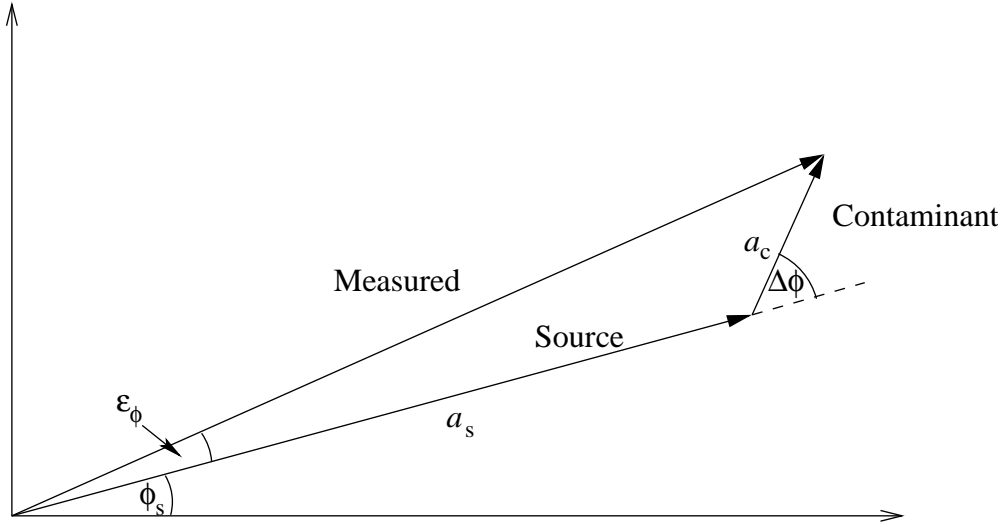}
  \caption[Vector representation of the summation of the fringes due to
    the target source and background
    contamination]{\label{fig:angles}Vector representation of the
    summation of the fringes due to the target source and background contamination.}
\end{figure}

We are interested in the phase error, $\varepsilon_\phi$, introduced
into the measured fringe by the contaminating spectrum. Since we are
only interested in the phase information, we ignore the offset term
$F_{\mathrm{s}}+F_\mathrm{c}$, and represent the varying term as a
vector summation, as shown in figure \ref{fig:angles}, where
$a_\mathrm{s}$ and $a_\mathrm{c}$ represent the source and contaminant
fringe amplitudes as before. The angle $\Delta\phi$ is the difference
between the source and contaminant fringe phases, $\Delta\phi =
\phi_\mathrm{c}-\phi_\mathrm{s}$. Using the sin and cosine rules for
triangles we can show
\begin{equation}\label{eqn:contamination1}
\frac{\sin \varepsilon_\phi}{a_\mathrm{c}} = \frac{\sin
  \Delta\phi}{\sqrt{a_\mathrm{s}^2+a_\mathrm{c}^2+2a_\mathrm{s}a_\mathrm{c}\cos \Delta\phi}}.
\end{equation}

First we consider the case that the two spectra are of {\em similar
  form} and {\em very close in velocity}, so that $\Delta\phi$ is small. Then,
\begin{equation}
\sin{\varepsilon_\phi} \approx \frac{a_\mathrm{c}\Delta\phi}{a_\mathrm{s}+a_\mathrm{c}}.
\end{equation}
If we assume the source and contaminant fringe visibilities are
approximately equal, so that $a_\mathrm{s}/F_{\mathrm{s}} \approx
a_\mathrm{c}/F_\mathrm{c}$, then $a_\mathrm{c}/a_\mathrm{s} \approx
F_\mathrm{c}/F_{\mathrm{s}}$, which is equal to the flux ratio of the
two fringes. If the contaminating fringe is much fainter than the
source, so that $F_{\mathrm{s}}\gg F_\mathrm{c}$, therefore $a_\mathrm{s}\gg
a_\mathrm{c}$, and hence $\varepsilon_\phi$ is small, then
\begin{equation}
\label{eqn:bginitial}
\sin \varepsilon_\phi \approx \varepsilon_\phi \approx \frac{a_\mathrm{c}}{a_\mathrm{s}}\Delta\phi =
\frac{F_\mathrm{c}}{F_\mathrm{s}}\Delta\phi.
\end{equation}
Still assuming that the two spectra
are of similar type and velocity, then all wavelength
channels will see approximately the same phase difference between source and
contaminant fringes, and this error will be
systematically close to the same across all channels. Therefore, the same result will be expected finally even
after averaging over all channels. Since $\Delta\phi$ is proportional to the difference in velocity,
$\Delta v$, between source and contaminant, then we can write the final
error in the measured velocity, $\varepsilon_v$, simply as:
\begin{equation}
\boxed{\varepsilon_v \, \approx \,
  \overline{(F_{\mathrm{c}}/F_{\mathrm{s}})} \Delta v 
  \, \approx \, \frac{\overline{F_{\mathrm{c}}}}{\overline{F_{\mathrm{s}}}}
 \Delta v,}
\label{eqn:bgcorrelated}
\end{equation}
where now $\overline{F_{\mathrm{c}}}$ and $\overline{F_{\mathrm{s}}}$
signify mean fluxes for the entire spectra, rather than for individual
channels. In other words, the systematic velocity error due to a faint
contaminant of similar spectral type that is closely matched in
velocity is simply the velocity difference scaled by the flux ratio of
the contaminant to the source fringe. (The approximation
made in dividing the means in the second form of this equation is good
to first order, and provides a very convenient way to quickly estimate
the errors. See appendix \ref{app:divisionapprox} for a derivation and
discussion of when it is more appropriate to use the first form of the
equation. The same approximation is also made use of several times
below.)

This relation does not hold well to arbitrarily large velocity
differences, however. From the geometry of figure \ref{fig:angles} it
can be seen that a worst case scenario is where the contaminant in all
channels is systematically offset by an amount such that the
background contaminant vector is perpendicular to the measured vector
(or, approximately, where $\Delta\phi=\pi/2$). In this case,
$\varepsilon_\phi \approx a_\mathrm{c}/a_\mathrm{s} \approx
F_\mathrm{c}/F_{\mathrm{s}}$, so that:
\begin{equation}
\boxed{\varepsilon_v \approx \Gamma\frac{\overline{F_\mathrm{c}}}{\overline{F_\mathrm{s}}},}
\label{eqn:bgworstcase}
\end{equation}
where $\Gamma$ is the phase/velocity scaling factor. Hence the `worst
case scenario' error, where the velocity offset between source and
contaminant is the worst possible and the two spectra are very close
in form, is again simply proportional to the contaminant-to-source
flux ratio.

In the limit that the spectra are completely {\em dissimilar}, or are
sufficiently separated in velocity space that overlapping features are
in no way correlated, then the phase errors will be randomly
distributed across all channels. Following again from equation
\ref{eqn:contamination1}, we once again assume $F_{\mathrm{s}} \gg
F_\mathrm{c}$ and $a_\mathrm{s} \gg a_\mathrm{c}$, which allows us to
neglect the $a_\mathrm{c}^2$ and $\cos\Delta\phi$ terms; and again
that on average $a_\mathrm{s}/F_{\mathrm{s}} \approx
a_\mathrm{c}/F_\mathrm{c} \Rightarrow a_\mathrm{c}/a_\mathrm{s}
\approx F_\mathrm{c}/F_{\mathrm{s}}$. Now, however, taking $\Delta\phi$ as
uniformly randomly distributed, we can find the rms value for the
phase error in one channel as:
\begin{equation}
\mathrm{\mbox{rms}}(\sin\varepsilon_\phi) \approx
\mathrm{\mbox{rms}}(\varepsilon_\phi) \approx
\mathrm{\mbox{rms}}\left
  (\frac{a_\mathrm{c}}{a_\mathrm{s}}\sin\Delta\phi\right ) =
\frac{F_\mathrm{c}}{F_{\mathrm{s}}}\frac{1}{\surd 2}
\end{equation}
Assuming an average over $n$ independent channels
gives a $1/\surd{n}$ reduction in the final error, so that for
uncorrelated spectra, we can expect a final velocity error of:
\begin{equation}
\boxed{\varepsilon_v \approx \frac{\Gamma}{\sqrt{2n}}\,\frac{\overline{F_\mathrm{c}}}{\overline{F_\mathrm{s}}},}
\label{eqn:bguncorrelated}
\end{equation}
where $\Gamma$ is the phase-velocity scale factor for the
instrument. The error is now independent of
differential velocity between source and contaminating spectra, since
the two spectra no longer bear any relation to each other (although
it may be expected to vary systematically on velocity difference
scales corresponding to the line widths).


\subsubsection{Application to Moonlight and Stellar Contamination}\label{sec:mooncontamination}
Equations \ref{eqn:bgcorrelated}, \ref{eqn:bgworstcase}, and
\ref{eqn:bguncorrelated} can be applied directly to estimate the
magnitude of the errors introduced by background scattered moonlight
contamination. As an example, a $3\arcsec$ fiber with a bright-time
sky background of 19$\vmag\,\mathrm{arcsec}^{-2}$ due to scattered
moonlight from the atmosphere gives a total of $16.9\vmag$ of sky
background. For a magnitude 12 star, this gives a
source-to-contamination flux ratio of about 90. In the worst case
scenario, from equation \ref{eqn:bgworstcase}, assuming
$\Gamma\sim3700\msrad$ (corresponding to a $7\mm$ delay), we find
$\varepsilon_v\approx 41\ms$. This will apply where the stellar
spectrum is similar to the moonlight spectrum (not uncommon, since
most targets are sun-like), and in the case where the velocity
difference between star and moonlight, $\Delta v$, is coincidentally
around $\sim6\kms$. At smaller velocity differences, the error will
scale roughly linearly as $\varepsilon_v = \Delta v/90$ up to this
point (equation \ref{eqn:bgcorrelated}). After that, it will improve
again as $\Delta\phi$ increases to $\pi$, where the phase error once
again approaches zero. As $\Delta\phi$ increases, the behavior is
likely to be somewhat oscillatory, with a period of ~$2\pi\Gamma =
2.3\times10^4\ms$ owing to the geometry of figure \ref{fig:angles},
decaying until $\Delta v$ is large enough that the two spectra are
completely uncorrelated. For $n=1000$ \emph{independent} wavelength
channels (i.e., 4000 pixel channels with an LSF $\sim4$ pixels wide),
the error should then approach equation \ref{eqn:bguncorrelated}, with
an rms value of around $\varepsilon_v\approx0.8\ms$. This should also
be the typical error size when the star and moon spectra are very
different in form.

\begin{figure*}[tbp]
  \epsscale{1.0}
  \plotone{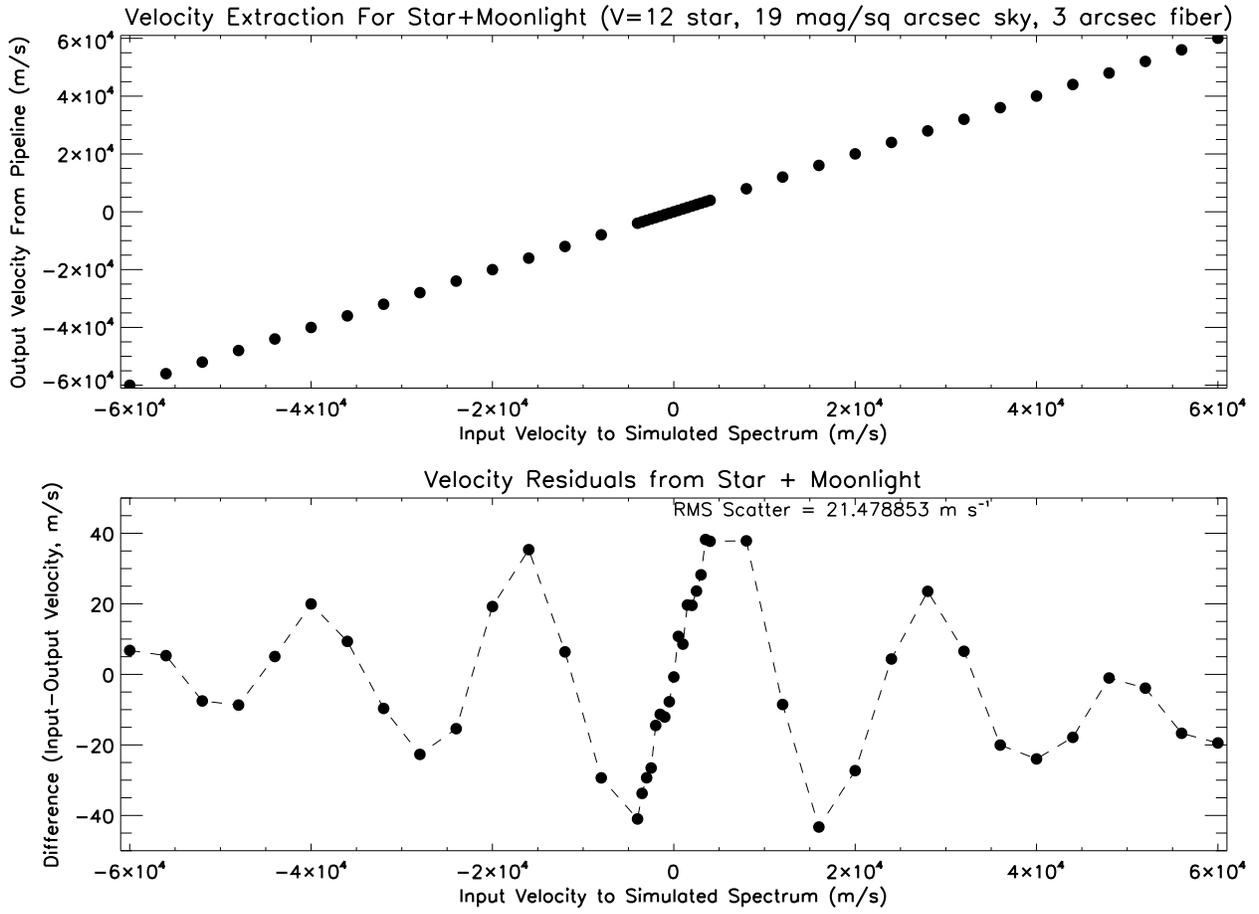}
  \caption[Simulations of moonlight contamination]{\label{fig:moonlightsim}Simulations of moonlight contamination, showing the
    systematic error introduced by contaminating moonlight at $19\vmag\,
    \mathrm{arcsec^{-2}}$ for a V=12 F9V star on a $3\arcsec$ fiber, as a function
    of velocity difference between target star and moon
    spectrum.}
\end{figure*}

Simulations of the effect of moonlight contamination show reasonable
agreement: figure \ref{fig:moonlightsim} shows the RV deviation caused
by synthetic moonlight contamination added to a synthetic stellar
spectrum, and then multiplied by the interferometer response function
and degraded in resolution to simulate real instrument spectra
(ignoring the iodine reference). The resulting spectra are run through
the standard ET reduction pipeline to assess the effects of the
contamination.

The instrument parameters given here are chosen to match the
parameters of the simulation, which reflect a typical ET-like
instrument design. For comparison, the MARVELS/Keck ET in fact uses
$1.8\arcsec$ fibers, with $\Gamma\sim3400\msrad$, leading to a
worst-case error of only $14\ms$, at the faintest end of the MARVELS
range. Reducing the fiber size is clearly one effective way of
mitigating the effect of moonlight contamination, although this may be
at the expense of throughput if telescope guiding or seeing is not
optimal. Reduction of moon contamination error is discussed further in
section \ref{sec:moonlightdiscussion}.

In exactly the same way, we can calculate the effects of
contamination by a background star: for example, a background star of
the same spectral type and class, but 5 magnitudes fainter
(i.e., fainter by a flux ratio of 100) would give about the same level
of error. At increasingly different spectral types, the contaminant
star will cause less of a problem as the spectra become less
correlated. Hence for a faint companion (as opposed to background)
star, although the flux ratio may be higher, the effect will be at
least partially mitigated by the difference in spectral type.


\subsection{Multiplicative Fringe Contamination}

In addition to additive contaminating spectra, certain errors can
appear as multiplicative effects in the fringing spectra. These are
independent of flux and correspond more closely to fringe errors
rather than flux errors in the spectra. Residual interferometer comb,
for example, will behave in this way (a concern for multiplied-reference
modes of operation), and the cross talk term from the reference
addition approximation can also be considered in the same way.

\subsubsection{Derivation}
We can follow the same formalism as for background spectrum
contamination. In this case, however, instead of the source and
contaminant fringe {\em visibilities} being similar, the fluxes are
similar, so that $F_{\mathrm{s}}\approx F_\mathrm{c}$. Dividing the
denominators of equation \ref{eqn:contamination1} through by
$F_{\mathrm{s}}\approx F_\mathrm{c}$, we can replace the fringe
amplitudes $a_\mathrm{s}$ and $a_\mathrm{c}$ with their respective
visibilities $\alpha_\mathrm{s} \equiv a_\mathrm{s}/F_{\mathrm{s}}$
and $\alpha_\mathrm{c} \equiv a_\mathrm{c}/F_\mathrm{c}$. The source
spectrum and interferometer comb are completely unrelated in
form. Assuming $\alpha_\mathrm{c} \ll \alpha_\mathrm{s}$ we can follow
the same reasoning as for equation \ref{eqn:bguncorrelated} and write:
\begin{equation}
  \boxed{\varepsilon_v \approx
    \frac{\Gamma}{\sqrt{2n}}\,\frac{\overline{\alpha_\mathrm{c}}}
    {\overline{\alpha_\mathrm{s}}},}
\label{eqn:combuncorrelated}
\end{equation}
where $\overline{\alpha_\mathrm{c}}$ and
$\overline{\alpha_\mathrm{s}}$ are representative visibilities for the
entire contaminant and source spectra respectively (again see appendix
\ref{app:divisionapprox} regarding the division of means here). The error is now
proportional to the ratio of visibilities, and again decreases with
the root of the number of spectral channels, $n$.

\subsubsection{Application to Residual Interferometer Comb}\label{sec:comberr}

As an example application, we can consider the effect of residual
interferometer comb. For the case of using a multiplied fiducial
spectrum such as in-beam iodine absorption, if the interferometer comb
(section \ref{sec:comb}) is not completely removed -- either by
careful tuning of interferometer delay and slit width or by Fourier
filtering in the data processing -- then it acts as contaminating
fringes. In order for the reference addition approximation to work,
the comb term in equation \ref{eqn:exactsolution} must be completely
removed (see the subsequent discussion). This has in the past been an
issue with some configurations of the ET instruments, for example: in
these cases the sampling by the resolution element was such that the
comb was aliased in places, creating a low frequency pattern in the
dispersion direction which was impossible to filter out without losing
significant Doppler information.

Here, it is appropriate to take the combined star/iodine data as the
source spectrum, since the comb error arises in the formula for the
combined data (equation \ref{eqn:exactsolution}).  A residual comb visibility of 0.5\% (c.f. $\sim1\%$ comb visibility in
unfiltered KPNO ET data) on top of a spectrum of typical mean
visibility of say, 4\%, and taking the KPNO ET value of
$\Gamma\sim3300\msrad$ with $n=1000$ independent channels, would give
an expected error of $\varepsilon_v \approx 9\ms$.

In practice, however, we have found that, provided comb aliasing is
avoided in the instrument design and alignment, removing the
interferometer comb during image preprocessing with a simple
one-dimensional low-pass Fourier filter appears to be effective in
completely mitigating this error. The comb cannot be measured or seen
by eye above the photon noise in filtered continuum lamp spectra, and
we have not yet found any evidence of residual comb causing problems
in the final data.


\subsubsection{Application to the Addition Approximation}\label{sec:additionapproxerr}

In order to estimate the errors introduced by the addition
approximation discussed in section \ref{sec:additionapprox}, we can
also follow a similar approach, treating the cross term from equation
\ref{eqn:exactsolution} which is
ignored in the approximation (or rather, treating the \emph{lack} of cross term)
as if it were a contaminating spectrum. First, we consider the simplified case of
two discrete overlapping Gaussian absorption lines, from template
spectra labeled A and B (e.g., an iodine and a stellar line), combined by
multiplication to give the measured spectrum, labeled M. Both line
centers are exactly coincident. The
fractional line depths are represented by $D$ ($0 \le D \le 1$), with
corresponding subscripts $a$, $b$ and $m$. From \cite{Ge2002}, we have
that in general:
\begin{equation}\label{eqn:ge2002} \alpha = De^{-3.56d^2/l_\mathrm{c}^2} \equiv KD, \end{equation}
where $\alpha$ is the absolute fringe visibility (so that the complex visibility is $\boldsymbol{\alpha} =
\alpha e^{i\phi}$ as usual), $d$ is the interferometer delay, $l_\mathrm{c} =
\lambda^2 /\Delta\lambda$ is
the coherence length of the interferometer beam with line width
$\Delta\lambda$ at wavelength $\lambda$, and $K =
\exp(-3.56d^2/l_\mathrm{c}^2)$ is a constant (for a given wavelength). Although not
very realistic, we begin by assuming both lines A and B and the
resulting line M are of similar width, and that the measured line,
which is the product of the two lines, is also 
approximately Gaussian. $K$ is then approximately the same for all
three lines. We can then write:
\begin{eqnarray} \label{eqn:sausages}
  \alpha_\mathrm{m} & \approx & D_\mathrm{m} K = [1-(1-D_\mathrm{a})(1-D_\mathrm{b})]K    \nonumber \\
  & = & [D_\mathrm{a} + D_\mathrm{b} - D_\mathrm{a} D_\mathrm{b}]K           \nonumber \\
  & = & \alpha_\mathrm{a} + \alpha_\mathrm{b} - D_\mathrm{a} D_\mathrm{b} K.
\end{eqnarray}

In the addition approximation, the complex visibilities of the
template spectra are added together. In this simple case, the two
lines are centered at the same wavelength and both are Gaussian, so
that one line is simply a scaled version of the other. By the
linearity of Fourier transforms, this means that the phases of the two
complex visibilities must be identical, so that in the addition
approximation, the two absolute visibilities add to give
$\alpha_\mathrm{m}\approx \alpha_\mathrm{a} + \alpha_\mathrm{b}$. The
remaining term in equation \ref{eqn:sausages}, is therefore
approximately the error, $\alpha_\mathrm{\varepsilon}$, the difference
between the added templates and the actual measured visibility:
\begin{equation}
\alpha_\mathrm{\varepsilon} = D_\mathrm{a} D_\mathrm{b} K
\end{equation} 

\begin{figure}[tbp]
    \epsscale{0.7}
    \plotone{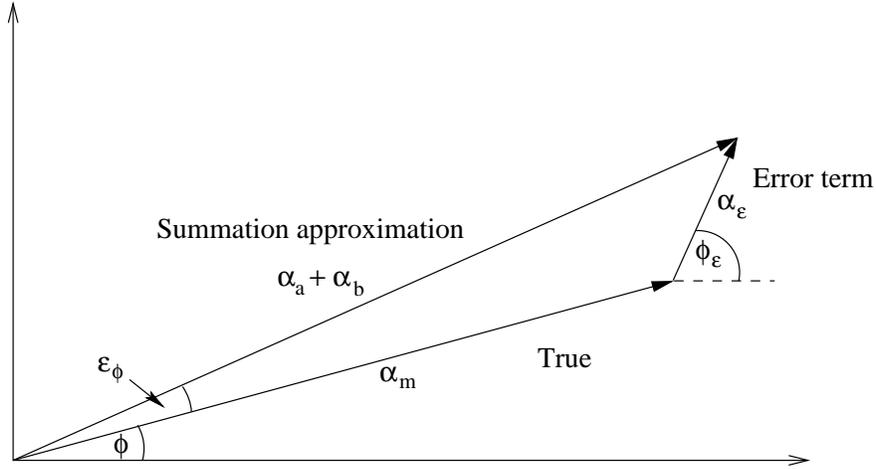}
    \caption[Vector representation of the summation of the true
    complex visibility and the error term due to the addition
    approximation]{\label{fig:AddApprxVectors}Vector representation of the summation of the true
    complex visibility and the error term due to the addition
    approximation.}
\end{figure}

In the more general case that the two line centers are not exactly
coincident or the same shape, so that the respective template fringes
are not in phase, the error term will also include a phase difference,
becoming a two dimensional vector, $\alpha_\mathrm{\varepsilon}
e^{\ii\phi_\mathrm{\varepsilon}}$. Taking the error term above as a
reasonable estimate of the length of this vector and assuming
$\phi_\mathrm{\varepsilon}$ is uniformly randomly distributed, we can
calculate a corresponding representative error in phase of the
summation approximation. Figure \ref{fig:AddApprxVectors} shows the
addition of the ``true'' (measured) complex visibility and the error
term to give the solution according to the summation
approximation. $\phi$ represents the phase of the true complex
visibility, and $\varepsilon_\phi$ represents the error in the
measurement of that phase. If we assume the resulting measured
visibility vectors and the error terms are uncorrelated from channel
to channel, and if we take $D_\mathrm{a}$ and $D_\mathrm{b}$ to be
some kind of representative average line depth for the two spectra
across all $j$, we can derive the typical expected velocity error
following the same reasoning as for equation
\ref{eqn:combuncorrelated} and write:
\begin{eqnarray}
  \varepsilon_v & = & \frac{\Gamma}{\sqrt{2n}} \, \frac{\alpha_\mathrm{\varepsilon}}{\alpha_\mathrm{m}}       \nonumber \\
             & = & \frac{\Gamma}{\sqrt{2n}} \,
             \frac{D_\mathrm{a}D_\mathrm{b}}{D_\mathrm{a}+D_\mathrm{b}-D_\mathrm{a}D_\mathrm{b}},
\label{eqn:additionerr}
\end{eqnarray}
where $n$ is again the number of independent channels, and the
constant $K$ cancels. (Note that although angles $\phi_\varepsilon$ in figure \ref{fig:AddApprxVectors} and $\Delta\phi$ in
figure \ref{fig:angles} are measured from different origins, they are
in both cases taken to be uniformly randomly distributed between
0 and $2\pi$, so that the same reasoning applies for both.)

Hence we find again that the error decreases as the square root of the
number of spectral channels; unsurprisingly, it also increases with
line depth, since deeper lines allow for more cross talk. If, however,
either one of the representative line depths is very small, then the
velocity error also becomes small, becoming approximately linear with
the smaller line depth and independent of the larger as the smaller
tends to zero.

\begin{figure}[tbp]
    \plotone{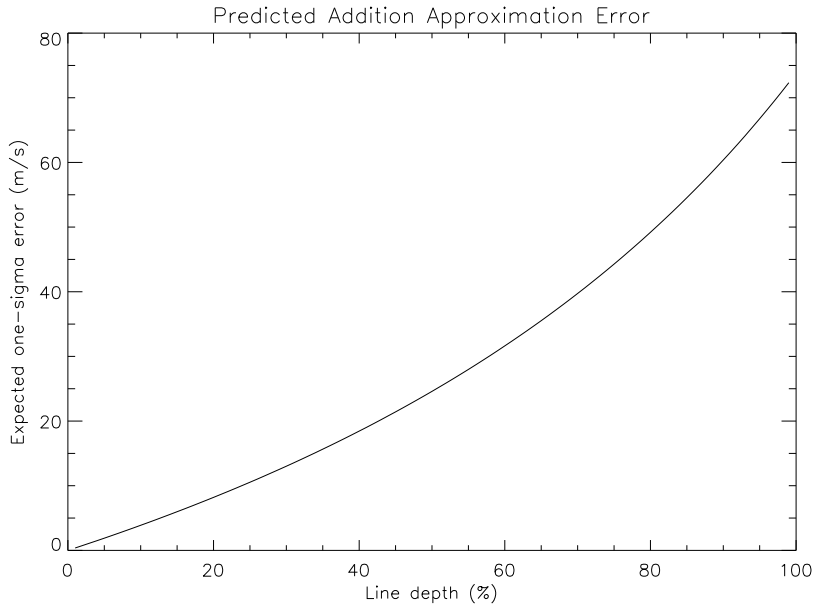}
    \caption[Analytically calculated expected error due to the addition
      approximation]{\label{fig:AdditionApproxErrTheory}Analytically
      calculated expected error due to the addition approximation,
      assuming approximately equal line depths for both star and
      reference spectra.}
\end{figure}

Figure \ref{fig:AdditionApproxErrTheory} shows the
expected typical error as a function of average line depth for the
simplified case where the typical depths of the two spectra are equal,
and taking $\Gamma \sim 3300\msrad$ and $n=1000$. For average
line depths of, say, 80\% for both star and iodine, this gives a
typical error due to the addition approximation of $\sim 50\ms$, which
is clearly very significant. The error will manifest as a systematic
error in the velocity response of the instrument, essentially adding
noise which varies as a function of the specific overlapping of the
lines between target star and reference spectrum. It will therefore
vary with stellar spectral type, class, and line width, and will also
vary as a function of the intrinsic absolute Doppler shift of the
stellar spectrum. Since the stellar lines are generally considerably
broader than the iodine lines, if the stellar lines slowly shift
relative to the iodine lines, the noise term will slowly change until
the point where a shift of more than a stellar line width has been
reached. At this point, the stellar lines are overlapping completely
new iodine features, and the noise term will take on a new value that
is completely uncorrelated with its previous value. Hence, we expect a
non-linearity in the velocity response of the instrument, with a
standard deviation somewhere on the order of $50\ms$ and that varies
with Doppler shift on a scale of approximately the line width of the
star. For solar-type stars observable with ET, this variation will be
over scales typically on the order of 5--$10\kms$.

\begin{figure*}[tbp]
  \epsscale{1.0}
  \plotone{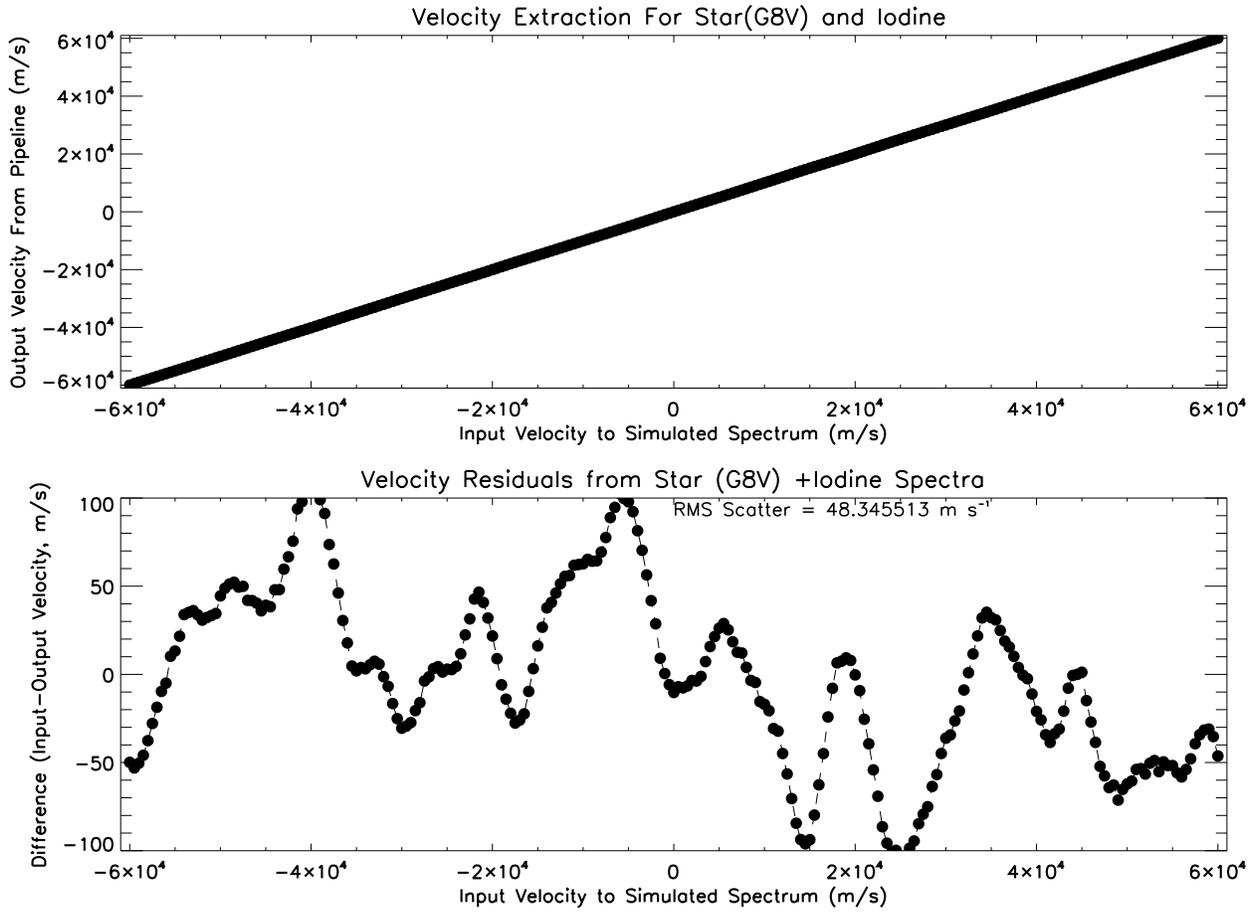}
  \caption[Simulations showing the addition approximation
  error]{\label{fig:SMnonlinearitysimulation}Simulations showing the
    addition approximation error. The non-linearity in the RV response
    has the same order of magnitude and appears on the same input
    velocity scales as expected from theoretical predictions.}
\end{figure*}

Figure \ref{fig:SMnonlinearitysimulation} shows the results of
simulated fringing spectra run through the reduction pipeline to see
the effect of non-linearity due to the addition approximation, and
shows broad agreement with these expectations. (For the simulation, the
phase-velocity scaling factor $\Gamma\approx 3700\ms$ was used; for this
value of $\Gamma$, our previous calculation yields $\sim
55\ms$.)

Clearly the addition approximation is a very significant source of
systematic error, and cannot be neglected. The systematic error will
affect any DFDI instrument that depends on in-beam multiplied
reference spectra. Various approaches to correct the error are under
consideration, although an exact analytical solution -- if one exists
-- remains elusive; for the MARVELS survey and the current KPNO ET
(now undergoing upgrade) simultaneous in-beam iodine calibration is
simply avoided, instead relying on good instrument stability and
bracketing exposures in time with reference iodine frames to calibrate
out instrument drift. Possible approaches to dealing with the addition
approximation error are discussed in section
\ref{sec:correctingadditionapprox}, and alternative calibration
methods that circumvent the approximation altogether in section
\ref{sec:CalAlternatives}.

%% file: section4.tex
\section{Discussion}\label{sec:discussion}

\subsection{Moonlight Contamination}\label{sec:moonlightdiscussion}
As surveying for exoplanets down to fainter and fainter magnitudes
during bright-sky time continues with the MARVELS project, moonlight
contamination is likely to become an important issue (keeping fiber
diameter small and avoiding bright time notwithstanding). With the
current ET instruments and pipeline, it is unlikely that direct
subtraction at the whirl or the initial image stage would be
successful. Bracketing science exposures in time with sky exposures to
measure the background would seriously impact the observing cadence,
reducing on-sky exposure time by a factor of two or more, and would
likely suffer from rapidly varying sky background in the presence of
even thin cloud. Using simultaneous sky-fibers alongside the science
fibers for direct subtraction in image space would require extremely
precise modeling of the instrument to successfully map the spectrum
from one fiber onto another. For subtraction in whirl space there is
insufficient flux from such a faint background to be able to
successfully measure meaningful whirls, at least using our current
data analysis techniques.

Given adequate templates of the moon spectrum, however (a solar
spectrum may suffice), then it may be possible to model out the moon
error by treating it as a further additive component in solving for
the stellar Doppler shift, just as for the case of added reference
spectra. This could in principle be done alongside any simultaneous
reference spectra, and since the moon spectrum is an additive
component, it would not suffer from the addition approximation errors
associated with an in-beam calibration source. Alternatively,
currently under consideration for reducing ET data, forward-modeling
from high-resolution spectra to match the measured whirl data (or even
the fringing spectrum images) could allow for moon contamination to be
included as a part of the model.

The original KPNO ET having been designed for brighter sources where
moonlight is less of a concern, approaches to mitigation of moonlight
contamination are still under investigation.

\subsection{The Addition Approximation
  Error}\label{sec:correctingadditionapprox}

One of the most significant concerns with the DFDI technique for
exoplanet searches when using superposed iodine is clearly the
addition approximation error. Causing long term systematic errors on the scale
of up to $\sim100\ms$, the approximation can potentially have a
serious adverse effect on the measurement of exoplanet RV
signatures. The effect can be mitigated to a certain extent simply by
judicious selection of observation times and positions of targets on
the sky, so that the line-of-sight barycentric motion of the Earth --
usually the dominant effect that causes the non-linearity to become
significant -- is minimized. Such observations are often not hard to
achieve at least over periods of a few days. This explains why we were
earlier still able to make successful detections of 51~Peg~b and
HD~102195~b \citep{51peg,ET1} despite then being unaware of the
effect: for both targets, the sky positions and epochs of observation
were such that the change in barycentric correction over the lengths
of the individual observing runs was small compared to the variation
scale of the addition error, so that the addition errors were absorbed
in small corrections to the phase-velocity scale. Nonetheless, the
addition approximation error becomes significant for velocity shifts
on scales upward of the line width of the stellar spectrum, and
placing stringent constraints on the times of observation is likely to
cause serious aliasing issues due to the observation window
function. Furthermore, over 24 hours, the barycentric motion of the observatory
contributes a variation of up to $\sim1\kms$ in radial velocity due to
the Earth's rotation, and up to $60\kms$ over a year due to the
Earth's orbital motion. For the slowest rotating, most narrow-lined
(and hence best Doppler precision) stars, the line width may be on the
order of $1\kms$, so that even over one night the error is of
concern. Forcing the observing cadence is therefore certainly not a
satisfactory long term solution.

Clearly, a robust solution to the problem of the addition
approximation error needs to be found. The ideal would be to find a
mathematically exact solution to the spectrum combination equations --
or at least a more accurate approximation -- but this remains elusive,
and it is not clear whether such a solution even exists. Modifying the
template whirls cannot solve the problem, since the templates
themselves are correct: to solve the problem, the cross-talk term
discussed in section \ref{sec:additionapprox} that constitutes the
error must be directly calculated, or at least approximated. Hence iterative
approaches which perturb the template whirls in an attempt to minimize
the residuals will only end up introducing error in order to fit the
cross talk term.

One way or another, calculating the cross-talk term seems to require
knowledge of the underlying high-resolution spectrum of the two
templates. Efforts to model the error term using high-resolution
iodine and synthetic stellar spectra have shown some promise. In this
approach, the cross talk term is calculated directly, so that a grid of
corrections across velocity and stellar parameter space can in
principle be created to apply to real data. Alternatively,
appropriately parametrized high resolution spectra could be forward
modeled to match the data from the instrument, using the formalisms
presented previously: as well as accounting for the cross talk term
naturally as part of the process, this could also allow for addition
of a third spectrum in the model to account for sky background in an
attempt to remove moonlight contamination. This would, however, likely
require extremely precise modeling and calibration of the instrument,
losing the benefit of the self-calibrating nature of real templates.

Alternatively, one can consider trying to obtain the high resolution
information from the data itself. Two possible approaches to
recreating the underlying spectrum are as follows: one is to begin
with the low-resolution non-fringing spectra obtained from the DFDI
fringing spectra (e.g., by binning the spectrum in the slit direction)
as an approximation, using this to help model the cross talk, and then
iterate with successive perturbations to the spectrum until the
residuals between real data whirl and the sum of the template whirls
and the cross-talk correction are minimized (similar to the approach
by \citet{JJohnsonNoTemplates} used to measure RV's without formal
templates using a traditional spectrograph). All the information
necessary to reconstruct the underlying spectrum may not be present in
the cross talk term, however: one can imagine degeneracies, for
example, where two closely spaced absorption lines within a resolution
element may lead to the same fringe phase and visibility as a single
line of a different depth positioned midway between the two. It
may therefore not be possible to iterate towards a single solution based on a
single data frame. However, once multiple observations at different
RV's have been taken, where the stellar lines overlap different parts
of the reference spectrum, we might conceivably be able to use the
aggregate information to help break any degeneracies. The more data
measurements, the more accurate becomes the estimate of the underlying
spectrum. This is somewhat analogous to the concept employed by
\citet{KonackiTatooine} in improving individual star templates from
double-lined spectroscopic binaries.

The second approach which is likely to be simpler is to try and obtain
an improved estimate of the cross-talk term by calculating it based on
template spectra reconstructed at higher resolution than the nominal
spectrograph resolution by using the information in the fringes. Such
a reconstruction is described by
\citet{ErskineResboost1Delay}. Although having a single fixed delay
constrains the degree to which high-resolution information can be recovered,
it could at least provide a first-order correction for the addition
approximation error.

At the time of writing, all these approaches represent avenues for further
investigation; the best solution may involve some combination of the above.

\subsection{Alternatives to Multiplied References} \label{sec:CalAlternatives}

Instead of attempting to calculate or model the addition approximation
error, an alternative is to circumvent the problem altogether by
considering different instrumental approaches to RV calibration.

One such approach would be `combined beam' superposition of the
reference spectrum, where a reference is literally added to the
stellar spectrum, for example, by splicing two input fibers together
into one. In this way, there is no longer an approximation in combining
the template whirls: the equations are exact (section
\ref{sec:addedref}). Combined-beam superposition does run the risk of
adding photon noise to the stellar spectrum: for an absorption
reference (e.g., a tungsten-illuminated iodine cell), the star and
reference spectra need to be balanced in flux. This adds
complexity to the observing in that it requires advance knowledge of
target fluxes and careful preparation for observations, and is more
likely to be practical for a single-object than a multi-object
instrument. Calculations of the added-reference photon limit based on
real KPNO ET fringing spectra suggest that provided the spectra are
properly balanced, similar precision can be obtained as for a
multiplied reference spectrum at a given S/N in the data, when one
allows for the gain in flux from the lack of reference
absorption. Similar test calculations with ThAr emission as a
reference show that the photon noise is less sensitive to the
relative intensity of an emission spectrum than an absorption spectrum
(see section \ref{sec:addedspecpherr}), relaxing the requirements on
intensity matching: such an approach may therefore overcome some of
the complexity of combined-beam observations and may even be practical
for a multi-object instrument.

Another intriguing question is whether the interferometer comb itself
could be used as a fiducial reference instead of iodine or
ThAr. Changes in the interferometer delay will shift the phase of the
comb, and so it can in principle track instrument drift, provided the
comb and star signals can be adequately separated. The problem lies in
the symmetry of the comb: as a simple example, if the image on the
detector were to drift in a direction exactly parallel to the comb,
the stellar fringes would appear to shift in phase and in the
dispersion direction, and yet the comb would appear not to have
changed at all, leading to the incorrect conclusion that the shift is
wholly intrinsic to the star. If the image on the detector can
reliably be stabilized to sufficient accuracy in either the slit or
the dispersion direction (or both), then there would be sufficient
information to break the degeneracy between stellar and instrument
shift, and the intrinsic RV could be measured. This would be a big
step forward, allowing simultaneous common-beam calibration with
neither flux loss (as in iodine absorption) nor photon noise addition
(as in ThAr superposition), and obviating the need for any reference
spectrum at all. (The USNO dFTS instrument \citep{USNOdFTS}, with its
lack of dependence on simultaneous in-beam calibration, is somewhat
similar in this respect, although a precise metrology system is needed
to measure the varying interferometer delay.) As pointed out by our
anonymous referee, the degeneracy could conceivably be broken given a
sufficiently large spectral bandpass: the wavelength dependence of the
comb frequency (see, e.g., figure \ref{fig:comb}) would allow for the
measurement of the image shift in the dispersion direction. The large
bandpass, however, would need to be balanced with the requirement that
the comb be resolved well enough to be measurable, which may be hard
with standard CCD detector sizes unless longer wavelengths are used
(since longer wavelengths exhibit a lower comb density in the
dispersion direction).

Finally, if instrument stability can be controlled well enough, simply
running parallel reference spectra alongside the stellar spectra, or
alternatively, bracketing stellar exposures in time with reference
exposures, can provide another solution: this has the twofold benefit
of vastly simplifying the data analysis and eliminating the
significant throughput loss ($\sim30$\%--50\%) due to insertion of an
iodine cell into the beam path. This is the current method of choice
for the MARVELS survey, and is also currently employed by the KPNO
ET. The MARVELS/Keck ET instrument is pressure stabilized, and
thermally controlled to the few-mK level, allowing for very good
instrument stability. Although not as precise as simultaneous common
path calibration, the results are adequate for the moderate-precision
large-scale survey for which the Keck ET is intended. On-sky results
with the Keck ET show that this approach is feasible, and exposure
bracketing is to be employed in the full survey.


\subsection{The Technique in Practice}\label{sec:inpractice}

\begin{figure}[tbp]
  \plotone{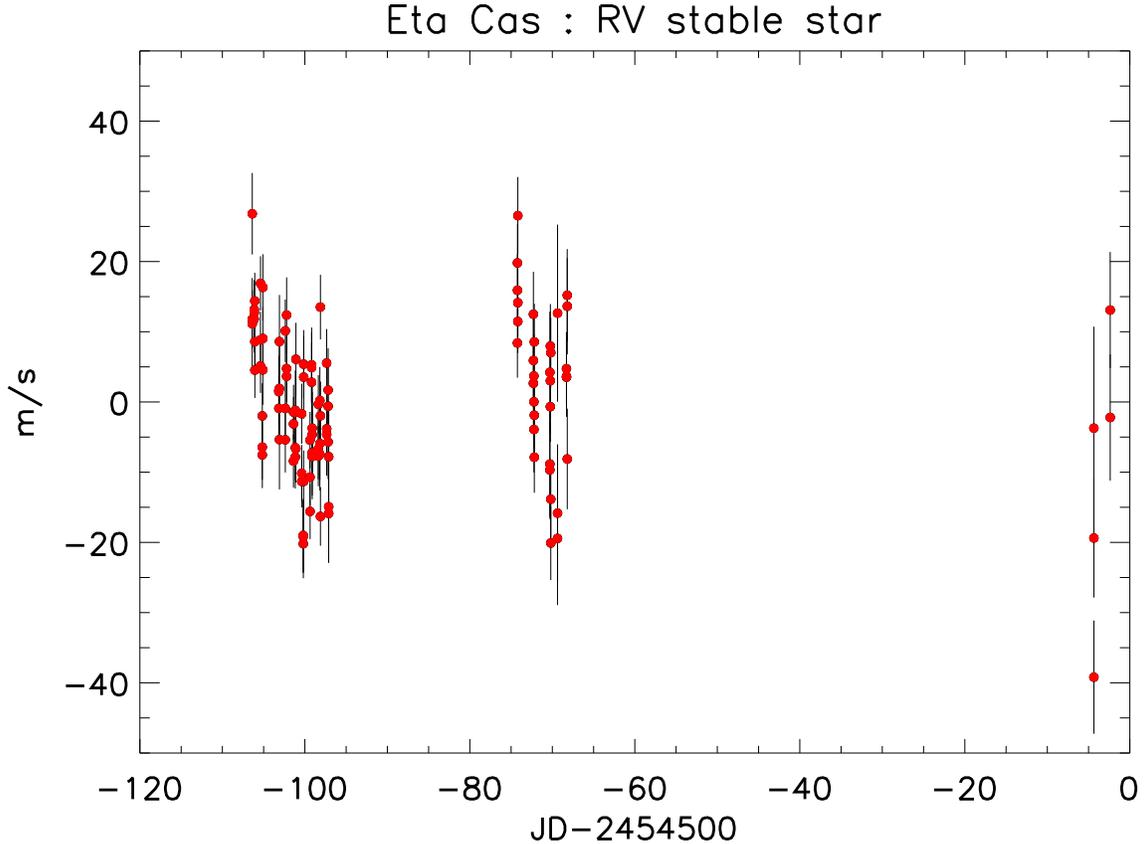}
  \caption{\label{fig:etacasplot}Differential RV
    measurements of $\eta$ Cas, an RV stable star, using the
    single-object KPNO ET. rms scatter is $10.8\ms$; error bars
    indicate the size of the photon error (mean $5.6\ms$), and do not
    include correlated systematic errors.}
\end{figure}

Beginning with the confirmation of 51~Peg~b, and with the later
discovery of HD~101195~b, the ET instruments have convincingly
demonstrated the capacity of dispersed fixed-delay interferometry for
exoplanet detection and discovery \citep{51peg,ET1}. Even in the
presence of the addition-approximation error, both the
single-object ET at KPNO and the multi-object Keck/MARVELS ET at APO
have been able to routinely uncover the RV signals of known exoplanets. The
early KPNO ET, using in-beam iodine, demonstrated photon-limited
precision at the 2--3$\ms$ level with bright reference stars on the
very short term (see earlier, figure \ref{fig:rmsvssn}). Figure
\ref{fig:etacasplot} shows observations of $\eta$~Cas, an RV-stable
star, using the same instrument over a longer period of several
months, with an rms of $10.8\ms$ (compared to a mean photon limit of
$5.6\ms$). Evidently in this case, the addition approximation error
due to the iodine was not too extreme. For comparison, typical rms
measurements on bright reference stars were on the order of 8--$10\ms$
over typical observing runs of $\sim 1$ week, where the addition
approximation error would normally be quite small (and to some extent 
absorbed in determination of the phase-to-velocity scale, $\Gamma$).

In anticipation of the upcoming MARVELS survey, the Keck instrument
saw a major upgrade in 2008, with a more stable mechanical design,
pressure stabilization, and extremely precise thermal control,
rendering the instrument stable enough to make exposure bracketing
with tungsten-illuminated iodine reference spectra feasible. As a
result, the addition approximation issue is eliminated and throughput
substantially increased (albeit at the expense of some loss of
precision due to the separated target and fiducial light paths). The
baseline requirements for the MARVELS survey with this approach call
for rms errors of $14\ms$ and $35\ms$ at $V=8\vmag$ and $V=12\vmag$
respectively, with corresponding photon error components of $3.5\ms$
and $21.5\ms$ \citep{GeMARVELSWhitePaper}. HD~9407 (stable,
$V=6.6\vmag$) shows an rms of $11.3\ms$ over four months, fairly
typical for stars at this brightest end of the target range; current
typical performance shows an rms of $15\ms$ at $V=8\vmag$, and $42\ms$
at $V=12\vmag$.

The KPNO ET is currently being upgraded, and in light of the addition
approximation error, observations are also now taken in an iodine
bracketing mode, with no simultaneous calibration. Further details and
results from both instruments can be found in
\citet{51peg,SPIE2004,ET1,suvrathsthesis,MyThesis,monolithic2008,SuvrathAbsoluteRV,GeMARVELSWhitePaper}. Such
precisions are adequate for finding planets with minimum masses
($M\sin i$) of order $1\Mj$ or more in few-day orbits
(i.e., hot Jupiters) down to $V=12$. They are also more than adequate
for uncovering stellar binary and brown dwarf companions.

We have presented here an overview of a mathematical basis for
understanding DFDI data for precision radial velocity measurements,
and discussed analytical approaches to some of the error sources that
would affect any implementation of the technique. The formulae derived
should prove useful for interpreting the data from any future
implementations of such instruments. As the ET instruments' overall
precision and reliability continues to improve, it is our hope that
the DFDI technique will be able to make a significant contribution to
the known extrasolar planet sample over the coming years.


%% file: acknowledgments.tex
The multiple-object ET instrument was supported by the W.~M.~Keck
Foundation, NSF with grant AST-0705139, NASA with grant NNX07AP14G,
the SDSS-III program, and the University of Florida. The development of
the single-object ET was supported by NSF with grant AST-0451407, NOAO,
and the Pennsylvania State University.

JvE and SM gratefully acknowledge support from the JPL Michelson
Fellowship program funded by NASA, and from Kitt Peak National
Observatory for travel support for many nights of observing at the
$2.1\m$ telescope. JvE also gratefully acknowledges support from the
fellowship provided by the Pennsylvania State University, and from the SPIE
scholarship program. This work has made use of the IDL Astronomy
User's Library \citep{IDlastrolib},\footnote{See also
\url{http://idlastro.gsfc.nasa.gov/}} and Craig Markwardt's IDL
library.\footnote{\url{http://cow.physics.wisc.edu/~craigm/idl/idl.html}}

Funding for SDSS-II has been provided by the Alfred~P.~Sloan
Foundation, the Participating Institutions, and the National Science
Foundation.

SDSS-III is managed by the Astrophysical Research Consortium for the
Participating Institutions. The Participating Institutions are the
University of Cambridge, Case Western Reserve University, University
of Portsmouth, Johns Hopkins University, Princeton University, the
Korean Scientist Group, the Max-Planck-Institute for Astrophysik, New
Mexico State University, Ohio State University, University of
Pittsburgh, Brazilian RIO Group, German Participation Group, New York
University, University of California-Santa Clara, University of
Florida, and the University of Washington.

The SDSS-III Web Site is: \url{http://www.sdss3.org}

The Center for Exoplanets and Habitable Worlds is supported by the
Pennsylvania State University, the Eberly College of Science, and the
Pennsylvania Space Grant Consortium.

We gratefully acknowledge our anonymous referee's input which helped
substantially improve this paper. J.vE. thanks Peter Plavchan for much
useful discussion of the addition approximation error, Dimitri Veras
for proofreading of early versions of the paper, and Justin Crepp for
his helpful suggestions.



{\it Facilities:} \facility{KPNO:2.1m (ET)}, \facility{Sloan (Keck ET)}, \facility{HET}.

%% file: appendixA.tex
\section{The Spectrograph Response Function and the LSF}\label{sec:AppLSF}

The response function $w_j(\lambda)$ due to the spectrograph optics is
very closely related to the instrument LSF. As
defined for the purposes of this paper, the spectrograph response
function specifies the respective instrumental throughputs for the
finite range of wavelengths, $\lambda$, falling at a given position in
the dispersion direction, $x=j$, on the detector, corresponding to a
spatially infinitesimally narrow channel in the spectrum. By contrast
the LSF specifies the flux distribution on the detector as a function
of spatial position in the dispersion direction due to a single
monochromatic wavelength of light. The response function at a
particular position is therefore determined by the way the LSF's from
all the different wavelengths overlap at that position.

If we assume the LSF is approximately identical in form at closely
separated channels $x$ on the detector, where $x$ represents the pixel
position in the dispersion direction (not necessarily integer), then
we can define the LSF as $L(x,x_0(\lambda_0)) = L_{\mathrm{t}}(x-x_0(\lambda_0))$
where $L$ represents the normalized envelope of flux spread
across pixels $x$ on the detector due to monochromatic light of
wavelength $\lambda_0$, centered at position $x_0$. $L$ is
simply a shifted version of $L_{\mathrm{t}}(x)$, which represents a template of
the LSF centered at $x = 0$. In general $x(\lambda)$ represents
the wavelength calibration mapping wavelength to detector position.

The response function at an infinitesimally wide position on the
detector, $w$, is given by writing the contribution from each overlapping LSF at that
position. The contribution from wavelength $\lambda_0$ at position $x_1$
is given by $w(\lambda_0,x_1) = L(x_1,x_0(\lambda_0)) = L_{\mathrm{t}}(x_1 -
x_0(\lambda_0))$. Therefore as a continuous function of general
wavelength $\lambda$, we can write the response function at position
$x_1$ as
\begin{equation}
w(\lambda,x_1) = L_{\mathrm{t}}(x_1 - x(\lambda)).
\end{equation}
We see that this is really just the LSF reversed (since the $x$ term is now negative).

We can now extend this to the total contribution across an entire
pixel at position $x=j$ where $j$ is now an integer representing pixel
number. Let $t_j(x^\prime)$ represent the pixel response function,
describing the normalized throughput of the pixel across its width as
a function of $x^\prime$. Then we can write
$w(\lambda,x^\prime)t_j(x^\prime)\,\mathrm{d} x^\prime =
L_{\mathrm{t}}(x^\prime -
x(\lambda))t_j(x^\prime)\,\mathrm{d}x^\prime$. Summing over all
$x^\prime$, we have the full response function $W_j$ for pixel column
$j$, given by:
\begin{equation}
W_j(\lambda) = \int L_{\mathrm{t}}(x^\prime - x(\lambda))t_j(x^\prime)\,\mathrm{d}x^\prime,
\end{equation}
i.e., essentially a convolution of the response function with the
pixel response function. To the extent that the width of the pixel is
narrow compared to the LSF (i.e., that the image is well over-sampled),
then to a reasonable approximation, $t_j$ is close to a delta
function, $W\approx w$, and the instrument response function at position $x$ is
approximately just the reversed LSF. Analogous arguments can be
followed in wavenumber ($\kappa$) space instead of wavelength space,
simply replacing $\lambda$ with $\kappa$ to obtain the same exactly
the same results as a function of $\kappa$.

%% file: appendixB.tex
\section{Fringe Formation: an Alternative Viewpoint} \label{sec:APanotherapproach}

We can view the formation of the fringes as given by equation
\ref{eqn:cplxvis} in another way. Beginning again with equation
\ref{eqn:howitallbegins} for the flux at a given channel, $j$, as a
function of delay $d$, and again substituting $Q$, we can write
\begin{eqnarray}\label{eqn:alternative1}
I_j(d)\; & = & \; \int Q(\kappa)\,\Re\{1+ \mathrm{e}^{-\ii 2\pi \kappa
  d}\}\,\dd \kappa \nonumber \\
 & = & \int P(\kappa)w_j(\kappa)[1+\cos(2\pi\kappa d)]\dd \kappa,
\end{eqnarray}
where again the spectrum within the channel is given by $Q(\kappa) =
P(\kappa) w_j(\kappa)$, with $P(\kappa)$ being the full spectrum
entering the instrument, and $w_j(\kappa)$ being the spectrograph
response function for channel $j$.

If we assume that the spectrograph response function is uniform across
the whole spectrum (i.e., for all $j$), then we can express it as a
wavenumber-shifted version of a global `template' spectrograph response function,
$w_{\mathrm{t}}$ (which is centered at $\kappa = 0$), shifted so that its center is at the
central wavenumber of the channel in question, $\kappa_j$. From
appendix \ref{sec:AppLSF}, we know that the spectrograph response
function is just the reverse of the LSF, so we can write
\begin{equation}
w_j = w_{\mathrm{t}}(\kappa-\kappa_j) = L_{\mathrm{t}}(\kappa_j-\kappa),
\end{equation}
where $L_{\mathrm{t}}$ represents a global template LSF, also centered at
$\kappa=0$. If we furthermore define $T(\kappa,d) \equiv
1+\cos(2\pi\kappa d)$, we can therefore rewrite equation
\ref{eqn:alternative1} as
\begin{equation}\label{eqn:alternative2}
I_j(d) = \int P(\kappa)T(\kappa,d)L_{\mathrm{t}}(\kappa_j-\kappa)\dd \kappa.
\end{equation}
$T$ can be thought of as the interferometer transmission function,
equivalent to the interferogram that would be obtained for pure white
light and an infinite resolution spectrograph (exactly as in equation
\ref{eqn:comb}). Equation \ref{eqn:alternative2} can be identified as
a convolution over the variable $\kappa$:
\begin{equation}
I_j(d,\kappa_j) = [P(\kappa)\, T(\kappa,d)]\otimes L_{\mathrm{t}}(\kappa),
\end{equation}
where the convolution is evaluated at wavenumber $\kappa_j$.

In other words, we have simply the input spectrum multiplied with the
interferometer transmission function, and then convolved with the LSF
due to the spectrograph. Thinking in two dimensions, to match the
wide-slit format of the actual ET spectra, we can replace the LSF with
its two-dimensional equivalent, the instrument point spread function
(PSF). Exactly the same results can be derived in frequency space,
simply by substituting frequency $\nu$ for $\kappa$ and time delay
$\tau$ for $d$.

This way of looking at fringe formation is the approach used by
\citet{Erskine03} and followed by \citet{suvrathsthesis}, and can
conveniently be employed to quickly produce simulated DFDI spectra
(although with the caveat that it assumes a uniform LSF, which in
practice is unlikely to be very realistic).

%% file: appendixC.tex
\section{Division-of-Means Approximation}\label{app:divisionapprox}

In section \ref{sec:contaminatingspectra} we make an approximation
regarding the magnitude of velocity errors resulting from a
contaminant spectrum, where we state that the error in velocity is
approximately equal to the flux ratio of the contaminant to the true
source spectrum multiplied by the velocity difference between the two
spectra (equation \ref{eqn:bgcorrelated}). Namely, we assume that
$\langle F_{\mathrm{c}}/F_{\mathrm{s}}\rangle \approx \langle
F_{\mathrm{c}}\rangle /\langle F_{\mathrm{s}} \rangle$ (we will use
$\langle\ldots\rangle$ to represent the mean here for notational
convenience). The same is assumed several times in the same section
(equations \ref{eqn:bgworstcase} and \ref{eqn:bguncorrelated}). This
approximation holds true provided that the fractional variation in the
power spectrum in the denominator with wavelength is predominantly
relatively small. We also make the same approximation regarding the
division of mean visibilities in sections \ref{sec:comberr}
and \ref{sec:additionapproxerr} (equations
\ref{eqn:combuncorrelated} and \ref{eqn:additionerr}). Here we show
the validity of this approximation.

Consider two arbitrary functions $A(x)$ and $B(x)$, which have
fractionally relatively small variations about their means so that we
can define two corresponding functions, $a(x)$ and $b(x)$ such that
\begin{equation}
A(x) \equiv \langle{A}\rangle(1+a(x)) \quad ; \quad
B(x) \equiv \langle{B}\rangle(1+b(x)),
\end{equation}
where $a,b\ll1$ for all $x$. Using the binomial expansion, we can write:
\begin{eqnarray}
\left\langle\frac{A}{B}\right\rangle
& = & \left\langle\frac{\langle{A}\rangle(1+a)}{\langle{B}\rangle(1+b)}\right\rangle
= \frac{\langle{A}\rangle}{\langle{B}\rangle}\left\langle\frac{1+a}{1+b}\right\rangle \nonumber \\
& = & \frac{\langle{A}\rangle}{\langle{B}\rangle}\left\langle(1+a)(1-b+b^2-b^3+\dots)\right\rangle \nonumber \\
& = & \frac{\langle{A}\rangle}{\langle{B}\rangle}\left\langle 1-b+b^2-b^3+\dots+a-ab+ab^2-ab^3+\dots\right\rangle \nonumber \\
& \approx & \frac{\langle{A}\rangle}{\langle{B}\rangle}\left[1+\langle b^2\rangle-\langle ab\rangle\right ],
\end{eqnarray}
where we have neglected terms of order $b^3$ and $ab^2$ and smaller,
and where we can also drop the terms $\langle a\rangle$ and $\langle
b\rangle$, which must equal zero according to the definitions of $a$
and $b$.

In general $\langle ab\rangle\rightarrow 0$ if the functions $a$ and
$b$ (and hence $A$ and $B$) are uncorrelated. In the limit where the
two functions are completely correlated (i.e., identical), then
$\langle ab\rangle\rightarrow \langle b^2\rangle$, and the $\langle
ab\rangle$ and $\langle b^2\rangle$ terms cancel, so that $\langle
A/B\rangle \rightarrow \langle A\rangle/\langle B\rangle$. In the
cases we are interested in, it is generally unlikely that $\langle
ab\rangle \ll 0$ (anticorrelation), or that $\langle ab\rangle \gg
\langle b^2\rangle$. Therefore, we can reasonably take $\langle
b^2\rangle$ as an estimate of the fractional error in the division
approximation. Conveniently, from the definition of $b$, $\langle
b^2\rangle$ turns out to be equal to the square of the normalized
standard deviation, $\sigma_B$, of the function $B$:
\begin{equation}
\langle b^2\rangle = \frac{\langle (B - \langle B\rangle)^2 \rangle}{\langle B\rangle^2} = \left(\frac{\sigma_B}{\langle B\rangle}\right )^2
\end{equation} 

Hence for functions with small fractional deviations from their
respective means, the mean of the quotient is approximately equal to
the quotient of the means, and $(\sigma_B/\langle B \rangle)^2$ gives
an estimate of the fractional error in the approximation.

Tests with stellar spectra show the approximation works quite well,
both at ET-like resolutions and with very high resolution synthetic
spectra. Taking as an example $R\sim 5000$ spectra from the KPNO ET
of 36~UMa (F8V) and $\tau$~Cet (G8V) (obtained by binning the fringing
spectra along the slit direction), we find $\sigma_{\mathrm{\tau
    Cet}}/\langle F_{\mathrm{\tau Cet}}\rangle = 0.25$, giving an
estimated error of 6\% due to the approximation. In practice, we find
the ratio of $\langle F_{\mathrm{36UMa}}\rangle/\langle
F_{\mathrm{\tau Cet}}\rangle$ to $\langle
F_{\mathrm{36UMa}}/F_{\mathrm{\tau Cet}}\rangle$ to be 0.980, i.e., a
2\% difference. For an {\em emission} spectrum in the denominator,
such as ThAr, the approximation will not hold well as there are
significant regions of near-zero flux. However, it is unlikely that
one would be interested in considering some contaminant spectrum
against a primary emission spectrum source. Conceivably one might be
interested in considering the effects of contamination {\em from} an
emission source -- e.g., from stray fluorescent lighting leakage, sky
emission, or leakage from an reference lamp, but in this case the
emission spectrum would be in the numerator: a stellar spectrum will
always be the function in the denominator, and so the approximation
should still hold.

The second case where the approximation is employed is in considering
visibility ratios, as in estimating the velocity error due to the
reference whirl addition approximation, or estimating the effect of
residual interferometer comb (equations \ref{eqn:combuncorrelated} and
\ref{eqn:additionerr}). In both cases here the function in the
denominator is absolute visibility as a function of wavelength,
$\alpha(\lambda)$, for reference-multiplied stellar data (since the
comb and cross-talk terms both appear in the formula for the combined
star/reference data, equation \ref{eqn:exactsolution}). Taking KPNO
spectra as an example, the normalized variance of the combined
star/iodine data tends to be around $(\sigma_\alpha/\langle \alpha
\rangle)^2\approx 0.32$, in other words giving an estimated error in
the division approximation of around 30\% for both comb and cross-talk
error sources -- not as precise as the approximation for contaminating
spectra, but still useful for an order-of-magnitude error. In neither
of these cases do we expect any correlation between the data
visibility function in the denominator and the comb or cross-talk
visibility function in the numerator, so the $\langle ab\rangle$ term
should be small.

Where there is concern over the assumptions made above (e.g., where the
normalized variance of the denominator, $(\sigma_B$/$\langle
B\rangle)^2$ is not small -- which would, for example, be the case
were the denominator to represent visibilities of a pure star
spectrum, where the visibility distribution peaks at very low
values), then the approximation derived here is not appropriate,
and instead it is necessary to calculate $\langle
F_{\mathrm{c}}/F_{\mathrm{s}}\rangle$ directly. Often this is in fact
entirely practical; however the approximation can generally be used to
give a very quick and convenient first-order estimate of contamination
errors.